\documentclass{aa}  

\usepackage{graphicx}
\usepackage{txfonts}
\def\nh3{NH$_{3}$}
\def\kms{km~s$^{-1}$}

\def\Vlsr{$V_{\rm LSR}$}
\def\Jyb{Jy~beam$^{-1}$}

\def\G24{G24.78$+$0.08}

\newcommand{\ms}{$M_{\odot}$}
\newcommand{\ls}{$L_{\odot}$}
\newcommand{\msyr}{$M_{\odot}$~yr$^{-1}$}
\newcommand{\pas}{$\rlap{.}^{\prime\prime}$}

\usepackage{color}

\begin{document}

   \title{A 10-\ms\ YSO with a Keplerian disk and a nonthermal radio jet}

   \subtitle{}
   
   \titlerunning{A Keplerian disk around a 10~\ms\ YSO}

   \author{L. Moscadelli \inst{1}
          \and
          A. Sanna \inst{2}   
          \and       
          R. Cesaroni  \inst{1}
          \and
           V.~M. Rivilla \inst{1}  
          \and 
          C. Goddi \inst{3,4}  
          \and
          K.~L.~J. Rygl \inst{5}     
          }

   \institute{INAF-Osservatorio Astrofisico di Arcetri, Largo E. Fermi 5, 50125 Firenze, Italy \\ 
              \email{mosca@arcetri.astro.it}
              \and 
             Max Planck Institut f\"ur Radioastronomie, Auf dem H\"ugel 69, 53121, Bonn, Germany 
             \and 
              Leiden Observatory, Leiden University, PO Box 9513, 2300 RA Leiden, The Netherlands
             \and
             Department of Astrophysics/IMAPP, Radboud University, PO Box 9010, 6500 GL, Nijmegen, The Netherlands
             \and
             INAF - Istituto di Radioastronomia \& Italian ALMA Regional Centre, Via P. Gobetti 101, 40129, Bologna, Italy}


 
  \abstract
   {To constrain present star formation models, we need to simultaneously establish the dynamical and physical properties of disks and jets around young stars.}
   {We previously observed the star-forming region G16.59$-$0.05 through interferometric observations of both thermal and maser lines, and identified a high-mass young stellar object (YSO) which is surrounded by an accretion disk and drives a nonthermal radio jet. Our goals are to establish the physical conditions of the environment hosting the high-mass YSO and to study the kinematics of the surrounding gas in detail.}
   {We performed high-angular-resolution (beam FWHM $\approx$~0\farcs15) 1.2-mm continuum and line observations towards  G16.59$-$0.05 with the Atacama Large Millimeter Array (ALMA).}
   {The main dust clump, with size  $\approx$10$^4$~au, is resolved into four distinct, relatively compact (diameter $\sim$2000~au) millimeter (mm) sources. The source harboring the high-mass YSO is the most prominent in molecular emission. By fitting the emission profiles of several unblended and optically thin transitions of CH$_3$OCH$_3$ and CH$_3$OH, we derived gas temperatures inside the mm~sources in the range \ 42--131~K, and calculated masses of \ 1--5~\ms. A well-defined Local Standard of Rest (LSR) velocity (\Vlsr) gradient is detected in most of the high-density molecular tracers at the position of the high-mass YSO, pinpointed by compact 22-GHz free-free emission. This gradient is oriented along a direction forming a large ($\approx$70\degr) angle with the radio jet, traced by elongated 13-GHz continuum emission. The butterfly-like shapes of the P-V plots and the linear pattern of the emission peaks of the molecular lines at high velocity confirm that this \Vlsr\ gradient is due to rotation of the gas in the disk surrounding the high-mass YSO. The disk radius is \ $\approx$500~au, and the \Vlsr\ distribution along the major axis of the disk is well reproduced by a Keplerian profile around a central mass of \ 10$\pm$2~\ms. The position of the YSO is offset by  \ $\gtrsim$0\farcs1 \  from the axis of the radio jet and the dust emission peak. 
To explain this displacement we argue that the high-mass YSO could have moved from the center of the parental mm~source owing to dynamical interaction with one  or more companions.
  }
{}


   \maketitle
%
\section{Introduction}

The formation of stars in the mass interval \ 1-20~\ms\ involves accretion disks and fast collimated outflows or jets. Atacama Large Millimeter Array (ALMA) observations at very high angular resolution ($\le$0\farcs1) have recently provided clear examples of disk-jet systems around young stellar objects (YSO) with luminosities  from low-mass to late O-type zero-age-main-sequence (ZAMS) stars \citep{Lee17,Gin18,San18}. From a theoretical point of view, the role played by turbulence, magnetic field, gravitational instability and radiation feedback in defining the  properties of the system (disk size and velocity profile; jet orientation and collimation) is still highly debated  \citep[][]{Mat17,Tan14,Kui18}. Detailed studies of disk-jet systems are therefore essential to constrain current star-formation models.

Since all star-formation models state that accretion and ejection are intimately related, studying the properties of the outflows from YSOs, which are much more extended and easier to observe than the small accretion disks, can be an effective way of probing the disk-jet systems.   Observations reveal different characteristics for solar-type and B-type YSOs: \ 1)~the molecular outflows tend to be less collimated with increasing YSO mass \citep{Beu05}, and \ 2)~only a few radio jets are known in B-type YSOs, whereas they are commonly observed towards low-mass protostars \citep[][]{Mos16,San18b}. If the larger distances and the more rapid evolution of B-type YSOs could in part explain these differences, we know that additional processes are at work in the formation of the most massive stars.
In comparison with low-mass protostars, the formation of B-type YSOs involves more energetic stellar radiation, which can ionize the surrounding gas and exert radiation pressure on it. The combined action of the magnetic field, the thermal pressure of the ionized gas, and the radiation pressure has been modeled by several authors \citep[see, e.g.,][]{Pet11,Vai11}, and results in a lower collimation of the outflow. Another notable difference is that more massive B-type stars form in richer clusters \citep{Car97,Hil98}, and dynamical and radiative interactions among cluster members can significantly affect the properties of the disk-jet systems, causing disk fragmentation, jet precession, or even  disruption of the system \citep[see, e.g.,][]{Far18}.  

In this paper we report on recent ALMA observations of the star-forming region (SFR) \ G16.59$-$0.05, also known as IRAS 18182$-$1433. This region has been the target of several millimeter (mm) and centimeter (cm) interferometric (Submillimeter Array, Owens Valley Radio Observatory, Plateau de Bure Interferometer, Very Large Array (VLA)) and Very Long Baseline Interferometry (VLBI) observations \citep{Beu06,Zap06,Fur08,San10a,Mos13b}. The bolometric luminosity of the region is \ $\sim$10$^4$~\ls\ \citep{Mos13b} \ at a distance of \ 3.6$\pm$0.3~kpc, as determined via maser trigonometric parallax observations \citep{Sat14}.  Inside a molecular clump of size~$\approx$0.5~pc and mass~$\approx$1900~\ms\ \citep{Beu02d}, the VLA C-Array observations at 3.6,~1.3,~and~0.7~cm by \citet{Zap06} identified two compact cm sources, named ``a'' and ``b'', separated by \ $\approx$2\arcsec\ along the SE--NW direction. From previous mm interferometric observations we know that, while source~``a'' coincides with the peak of the dust emission, indicating that this source is the most embedded, source~``b'' is found in good positional correspondence with intense, high-density molecular tracers, marking a hot molecular core (HMC).

European VLBI Network (EVN) observations of the 6.7-GHz methanol masers demonstrate that this maser emission is associated with the HMC and
traces an elongated structure of $\approx$2000~au, with the three-dimensional (3D) maser velocity pattern suggesting rotation \citep[][see their Fig.~7]{San10a} about a central mass of \ $\approx$12~\ms\ \citep{Mos13b}.
The sensitive Jansky Very Large Array (JVLA) A-Array observations at 5,~2.3,~and~1.3~cm by \citet[][see their Fig.~1]{Mos13b} resolve the compact cm~source~``b'' into a radio jet, elongated \ $\approx$3\arcsec\ along the E--W direction. The spectral index of the jet is negative, indicating nonthermal emission over
most of the jet, except the peak close to the maser disk, where thermal free-free emission is observed \citep[][see their Fig.~2]{Mos13b}. Water masers, monitored with the Very Long Baseline Array (VLBA) by \citet{San10a}, are distributed close to the HMC, and appear to trace a wide, fast bow-shock at the head of the western lobe of the radio jet. 

The combination of a rotating disk, as suggested by the 6.7-GHz maser proper motions, and a nonthermal radio jet, one of the few observed in high-mass SFRs \citep[see, e.g.,][]{Rod17}, makes \ G16.59$-$0.05 \ a promising target to constrain the formation mechanism of B-type stars. For this reason, we conducted new ALMA observations towards \ G16.59$-$0.05, as described in Sect.~\ref{obse}. In Sect.~\ref{resu}, we report our observational results presenting a general view of the physical conditions and kinematics of the gas at the center of the mm~clump. Section~\ref{disk-pro} derives the  properties of the disk and the YSO, and 
 Sect.~\ref{YDJ} provides a more general discussion of the geometrical and physical relations among the YSO, the disk, and the jet. Finally, our conclusions are presented in Sect.~\ref{conc}.

\section{ALMA observations}
\label{obse}

ALMA observed G16.59$-$0.05 during Cycle~3 in September 2016. Fourty-seven 12-m antennae were used in an extended configuration with baselines ranging from 15 to 3300~m. The observing time on the target source was about 1.8~h. The bandpass calibrator was the strong quasar \ J1924$-$2914, and the phases were calibrated from interleaved observations (every 8-9~min) of the quasar \ J1832$-$2039, separated from the target on the sky by $\approx$7\degr. To check the astrometric accuracy, ALMA also observed the quasar \ J1830-1606, separated from the target by $\approx$3\degr. We have verified that the peak emission of the phase-calibrated image of \ J1830-1606 \ differs from the nominal position by less than 10~mas, which provides an estimate of the astrometric uncertainty of our ALMA images.

The correlator frequency setup consisted of six spectral windows (SPWs), one broad 2-GHz spectral unit to obtain a sensitive continuum measurement at \ $\approx$242~GHz, and five narrower SPWs to cover a large number of lines, in particular the CH$_3$CN~(J~=~14-13), CH$_3$OH, and SiO rotational transitions. For each SPW, Table~\ref{cor-set} reports the frequency coverage, spectral resolution, and sensitivity, and also indicates the most prominent molecular species.

Data calibration was performed using the pipeline for ALMA data analysis in the Common Astronomy Software Applications \citep[\textsc{CASA},][]{McM07} package, version~4.7. For each SPW, looking at the plots of  (baseline-averaged) $uv$-amplitude versus channels, we selected the most intense emission line (always \ $\ga$10 times stronger than the continuum level of the $uv$-amplitude spectra), self-calibrated its channel-averaged emission, and, before imaging, applied the self-calibration phase solutions to all the channels of the SPW. Self-calibration improves the signal to noise ratio (S/N) of the final images of the SPWs by a factor of \ 1.1--1.5, with the larger gain factors obtained in correspondence with the stronger self-calibrated lines. The dynamical range of the line images varies in the range 20--50, depending on the considered SPW. The images for the continuum and line emissions were produced manually using the \textsc{CLEAN} task, with the robust parameter of \citet{Bri95} set to 0.5, as a compromise between resolution and sensitivity to extended emission. The clean beam FWHM of the resulting images varies in the range \ 0\farcs13--0\farcs17. The mm continuum image of \ G16.59$-$0.05 \ has a \ 1$\sigma$ rms noise level of \ 0.08~m\Jyb, limited by the dynamic range. The  1$\sigma$ rms noise in a single spectral channel varies in the interval \ 1--3~m\Jyb, depending on the considered SPW/channel.

We have employed a specific procedure to  determine the continuum level of the spectra and subtract it from the line emission. Since 
all the observed SPWs present a ``line forest'', identifying the channels where no line emission is present is a very difficult task. For each SPW, we use STATCONT\footnote{\url{http://www.astro.uni-koeln.de./~sanchez/statcont}} \citep{Sanc17}, a statistical method to estimate the continuum level at each position of the map from the spectral distribution of the intensity at that position. 

%
\begin{table*}
\caption{Spectral windows covered by the correlator set-up, with corresponding
spectral resolutions and noise per channel.}             
\label{cor-set}      
\centering                          
\begin{tabular}{c c c c c}        
\hline\hline                 
SPW & Frequency range & Resolution & 1-$\sigma$ noise$^a$  & Molecule$^{b}$ \\
 & (GHz) & (\kms) & m\Jyb &  \\    
\hline                        
 0 & 240.748--242.748  & 19 & 0.08  &  \\      
 1 & 260.151--260.619  & 0.56   & 1.8   &  SiO \\
 2 & 256.703--257.641  & 0.28   & 1.9   &   CH$_3$CN \\
 3 & 241.629--241.863  & 0.30   & 2.9   &  CH$_3$OH \\
 4 & 240.899--241.133  & 0.30   & 2.5   &  C$^{34}$S \\ 
 5 & 241.499--241.733  & 0.60   & 2.6   &  SO$_2$ \\ 
\hline                                   
\end{tabular}
\tablefoot{
\\ \tablefoottext{a}{For the SPW~0, the reported noise is for the frequency-averaged image, while for the other SPWs it has been estimated in the channel corresponding to the strongest emission in the bandwidth. This value is only indicative, as it can change significantly from channel to channel.}
\\ \tablefoottext{b}{Most prominent species in the SPW.}
}
\end{table*}
%

%
\begin{table}
\caption{List of the molecular transitions considered in this work. The transitions given in boldface characters have been fitted with \textsc{MADCUBA} to determine the gas physical conditions.}             
\label{used_lines}      
\centering                          
\begin{tabular}{c c c r}        
\hline\hline                 
Mol. Species & Frequency & Resolved QNs & $E_u$/$k$ \\    
            &  (MHz) &    &   (K) \\
\hline                        
 CH$_3$OH        &   241590.115  & 25$_{3,22}$--25$_{2,23}$ &  804 \\
                &  {\bf 241700.219 }  & {\bf 5$_{0,5}$--4$_{0,4}$ } & {\bf 48 } \\
                 & {\bf  241767.224 } &  {\bf 5$_{-1,5}$--4$_{-1,4}$ } & {\bf 40 } \\                 
                 & {\bf 241791.431 }  & {\bf  5$_{0,5}$--4$_{0,4}$ } & {\bf 35 } \\                
                 &  241806.508 &  5$_{4,1}$--4$_{4,0}$ & 115 \\
                 & 241829.646 & 5$_{4,1}$--4$_{4,0}$ & 131 \\  
                 & {\bf 241843.608 } & {\bf 5$_{3,2}$--4$_{3,1}$ } & {\bf 83 } \\            
                 & 257402.190  &  18$_{3,16}$--18$_{2,17}$  & 447 \\
                & 260381.560  &  20$_{3,18}$--20$_{2,19}$  & 537 \\

\hline
CH$_3$CN         & 257127.035 & J$_K$ =  14$_9$--13$_9$ & 671 \\     
                 & 257210.877 & J$_K$ =  14$_8$--13$_8$ & 549 \\  
                 & 257284.935 & J$_K$ =  14$_7$--13$_7$ & 442 \\
                 & 257349.179 & J$_K$ =  14$_6$--13$_6$ & 350 \\
                 & 257448.128 & J$_K$ =  14$_4$--13$_4$ & 207 \\
                 & 257482.791 & J$_K$ =  14$_3$--13$_3$ & 157 \\
                 & 257507.561 & J$_K$ =  14$_2$--13$_2$ & 121 \\
\hline
C$^{34}$S        & 241016.089 & J =  5--4   &  28 \\
\hline
{\bf CH$_3$OCH$_3$ } & {\bf 240978.322 } & {\bf 5$_{3,3}$--4$_{2,2}$ } & {\bf 26 } \\
               & {\bf 240982.799  } & {\bf 5$_{3,3}$--4$_{2,2}$  } & {\bf 26  } \\
               & {\bf 240985.078  } & {\bf 5$_{3,3}$--4$_{2,2}$  } & {\bf 26  } \\
               & {\bf 240989.939  } & {\bf 5$_{3,3}$--4$_{2,2}$  } & {\bf 26  } \\
               & {\bf 241523.829  } & {\bf 5$_{3,2}$--4$_{2,3}$  } & {\bf 26  } \\
               & {\bf 241528.306  } & {\bf 5$_{3,2}$--4$_{2,3}$  } & {\bf 26  } \\
               & {\bf 241528.692  } & {\bf 5$_{3,2}$--4$_{2,3}$  } & {\bf 26  } \\
               & {\bf 241530.972  } & {\bf 5$_{3,2}$--4$_{2,3}$  } &  {\bf 26 } \\
               & {\bf 241635.940  } & {\bf 21$_{3,18}$--20$_{4,17}$  } & {\bf 226  } \\
               & {\bf 241637.327  } & {\bf 21$_{3,18}$--20$_{4,17}$  } & {\bf 226  } \\
               & {\bf 241638.713  } & {\bf 21$_{3,18}$--20$_{4,17}$  } & {\bf 226  } \\
               & {\bf 241638.717  } & {\bf 21$_{3,18}$--20$_{4,17}$  } & {\bf 226  } \\
               & {\bf 257048.633  } & {\bf 18$_{2,16}$--17$_{3,15}$  } & {\bf  164 } \\
               & {\bf 257049.945  } & {\bf 18$_{2,16}$--17$_{3,15}$  } & {\bf  164 } \\
               & {\bf 257051.256  } & {\bf 18$_{2,16}$--17$_{3,15}$  } & {\bf  164 } \\
               & {\bf 257051.258  } & {\bf 18$_{2,16}$--17$_{3,15}$  } & {\bf  164 } \\
               & {\bf 260327.165  } & {\bf 19$_{5,15}$--19$_{4,16}$  } & {\bf  208 } \\
               & {\bf 260327.238  } & {\bf 19$_{5,15}$--19$_{4,16}$  } & {\bf  208 } \\
               & {\bf 260329.312  } & {\bf 19$_{5,15}$--19$_{4,16}$  } & {\bf  208 } \\
               & {\bf 260331.422  } & {\bf 19$_{5,15}$--19$_{4,16}$  } & {\bf  208 } \\
               & {\bf 260400.617  } & {\bf 16$_{5,11}$--16$_{4,12}$  } & {\bf  159 } \\
               & {\bf 260401.254  } & {\bf 16$_{5,11}$--16$_{4,12}$  } & {\bf  159 } \\
               & {\bf 260403.392  } & {\bf 16$_{5,11}$--16$_{4,12}$  } & {\bf  159 }  \\
               & {\bf 260405.845  } & {\bf 16$_{5,11}$--16$_{4,12}$  } & {\bf  159 } \\
               & {\bf 260615.883  } & {\bf 25$_{5,21}$--25$_{4,22}$  } & {\bf  332 } \\
               & {\bf 260615.883  } & {\bf 25$_{5,21}$--25$_{4,22}$  } & {\bf  332 } \\
               & {\bf 260616.584  } & {\bf 25$_{5,21}$--25$_{4,22}$  } & {\bf  332 } \\
               & {\bf 260617.285  } & {\bf 25$_{5,21}$--25$_{4,22}$  } & {\bf  332 } \\       
\hline
\end{tabular}
\end{table}
%

%
\begin{table*}
\caption{Physical parameters of the mm sources}             
\label{cont-sour}      
\centering                          
\begin{tabular}{c r c c c c c}        
\hline\hline                 
mm~source &   Flux & $\log$(N$_{\rm col}$)  & T$_{\rm ex}$  & \Vlsr  & FWHM   &  Mass \\
          &   (mJy)     &  \; \; \; (cm$^{-2}$) & (K) & (\kms) & (\kms) & (\ms) \\    
\hline                        
 B & 70$\pm$6 & 17.8$\pm$16.6 & 131$\pm$11 & 62.3$\pm$0.2 & 5.1$\pm$0.5 & 1.8$\pm$0.3 \\        
 A1 & 164$\pm$18 & 17.2$\pm$16.4 & 111$\pm$18 & 62.4$\pm$0.5 & 4$\pm$1  &  5.1$\pm$1.5 \\  
 A2 & 44$\pm$4 & 17.6$\pm$16.6 & 106$\pm$16 & 61.1$\pm$0.4 & 4.6$\pm$0.9 &   1.4$\pm$0.4  \\  
 A3 & 41$\pm$5 & 15.6$\pm$15.1 &  42$\pm$7  & 62.1$\pm$0.5 & 3.4$\pm$0.9 & 4.0$\pm$1.3  \\   
\hline                                   
\end{tabular}
\tablefoot{Column~1 indicates the mm source; column~2 lists the 1.2-mm flux calculated by integrating inside the contour at 10\% of the peak around each source (see Fig.~\ref{cont}); column~3 reports the column density of the molecular species fitted with \textsc{MADCUBA}: \ (1)~CH$_3$OCH$_3$ \ for sources B, A1, and A2, and (2)~CH$_3$OH for A3; columns~4,~5,~and~6 give, respectively, the excitation temperature, velocity, and line width of the fitted molecular lines; column~6 lists the estimated mass of the mm source. 
}
\end{table*}
%

\section{Results}

\label{resu}

This section presents the main results of our ALMA observations, first reporting on the physical  properties of the parental core and then focusing on the kinematics of the embedded high-mass YSO, the main target of our study.
 
\subsection{G16.59$-$0.05~B: the prominent molecular source in a cluster}
\label{res_stru}

\begin{figure*}
\centering
\includegraphics[width=\textwidth]{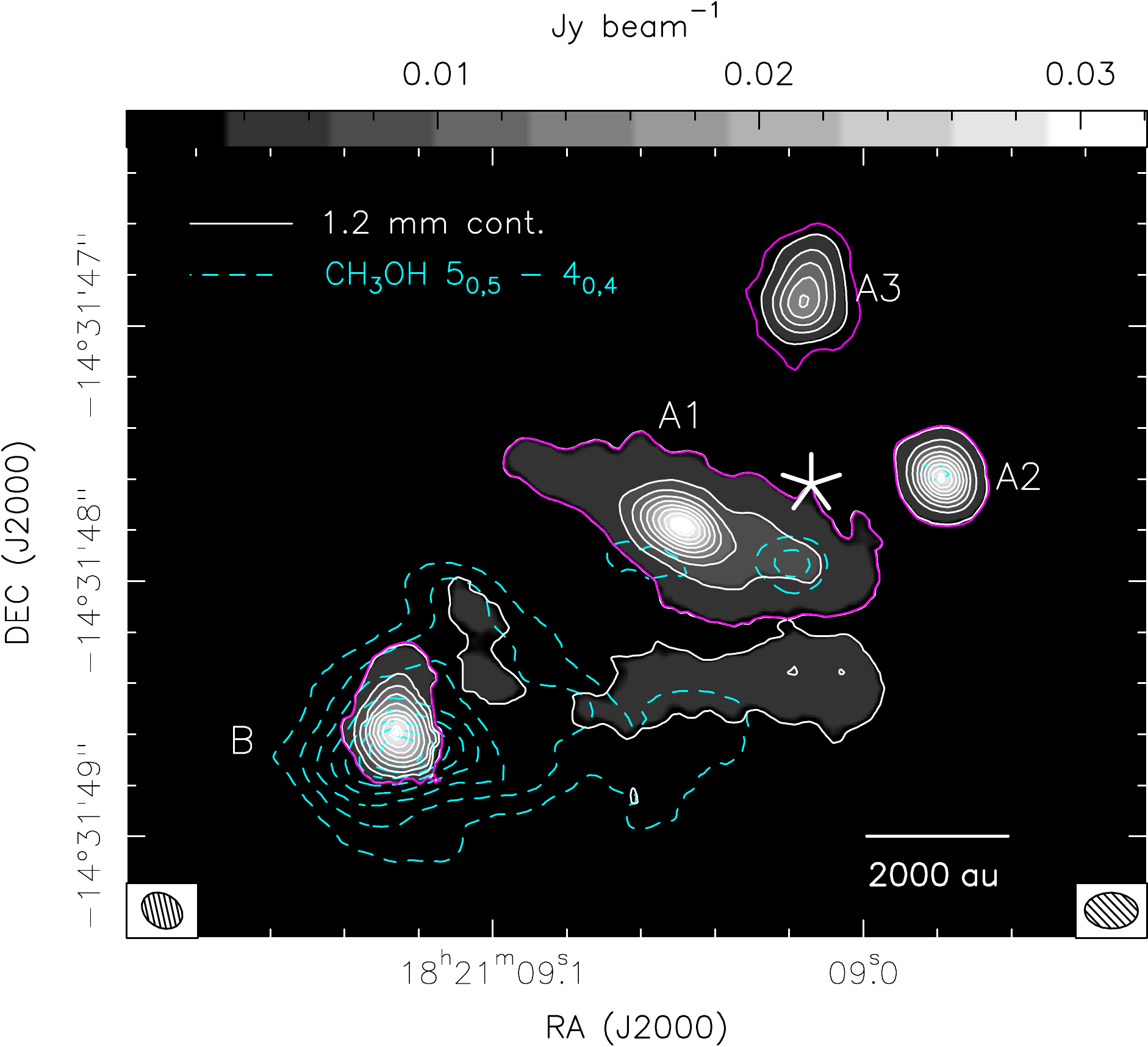}
\caption{The gray-scale map and white contours reproduce the ALMA 1.2-mm continuum. The color scale at the top gives the intensity of the map. The plotted levels are from 10 to 90\% of \ 0.032~\Jyb \ in steps of 10\%. The main continuum sources are indicated with their corresponding labels. The level at 10\% of the peak around each source is plotted in magenta. The velocity-integrated intensity of the \ CH$_3$OH 5$_{0,5}$--4$_{0,4}$ line is shown with dashed cyan contours, plotting levels from 20 to 90\% of 1.1~\Jyb~\kms\ in steps of 10\%. The ALMA beams for the continuum and the \ CH$_3$OH 5$_{0,5}$--4$_{0,4}$ line are reported on the bottom left and right corners, respectively. The big white star marks the position of source~``a'' \citep{Mos13b}.}
\label{cont}
\end{figure*}
The ALMA 1.2-mm continuum map presented in Fig.~\ref{cont} reveals four main sources, distributed from SE to NW over a region of $\sim$10$^4$~au. The source~B corresponds to the HMC identified in previous interferometric observations at 3~and~1~mm  \citep{Fur08,Beu06}, named source ``b'' in \citet{Mos13b}. The compact ``mm core'' from previous, lower-angular-resolution, interferometric observations, named source ``a'' in \citet{Mos13b}, is now resolved into three distinct sources, labeled A1, A2, and A3, separated by 2000--3000~au.  Considering emission down to 10\% of the peak, the sources B, A2, and A3 are relatively compact with size \ $\la$0\farcs5, while A1 has a significantly flatter spatial distribution extending up to \ $\approx$1\farcs5.


\begin{figure*}
\centering
\includegraphics[width=0.75\textwidth]{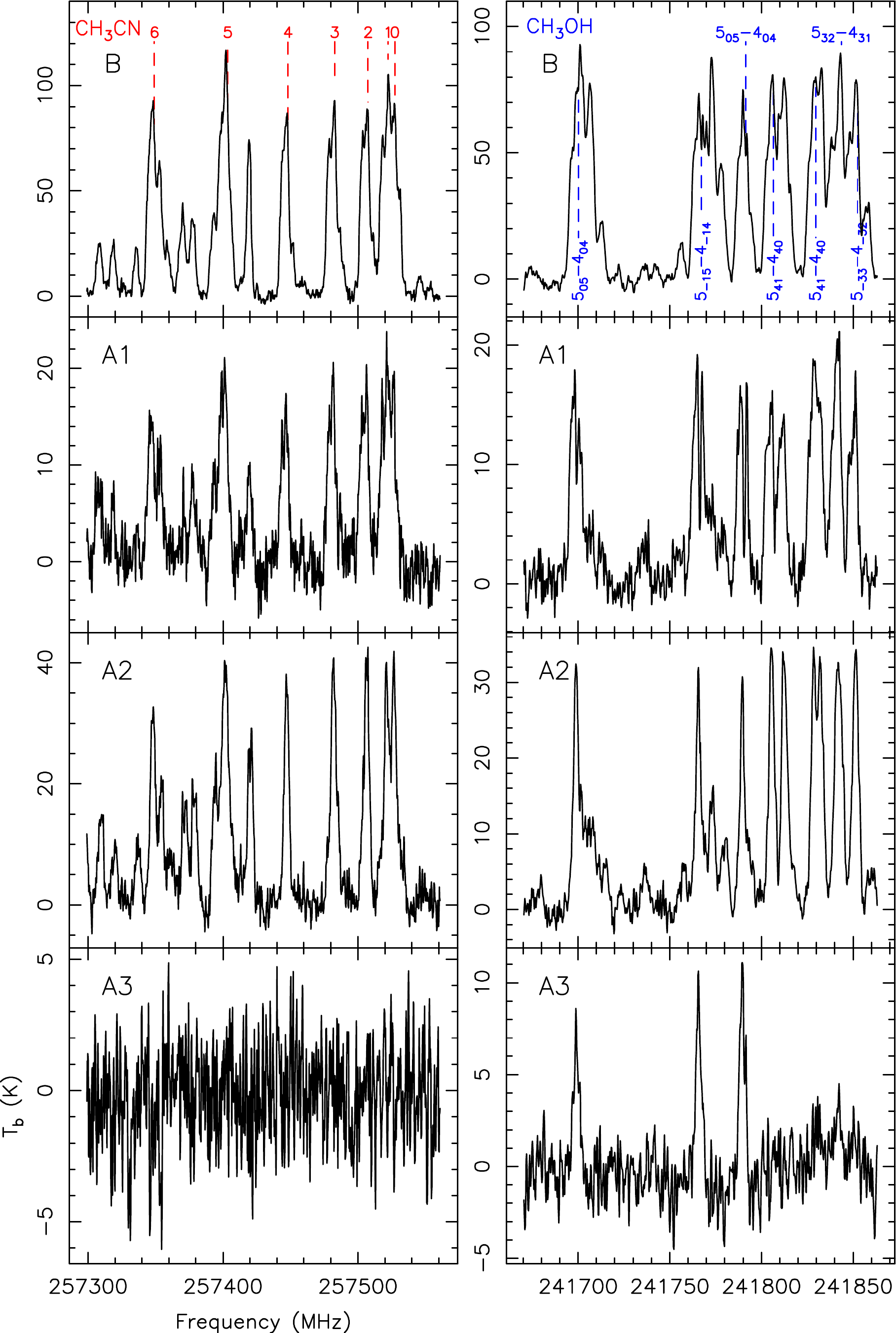}
\caption{Spectra of the prominent lines of \ CH$_3$CN~(J~=~14-13) ({\em left~panels}) and \ CH$_3$OH ({\em right~panels}) extracted at the position of the 1.2-mm peak for each of the four continuum sources \ B, A1, A2 and A3 (from top to bottom, respectively). The spectra are shown as brightness temperature vs. rest frequency. The transitions of the two molecular species (see Table~\ref{used_lines}) are labeled in the upper panels.}
\label{cont-spe}
\end{figure*}

\begin{figure*}
\centering
\includegraphics[width=0.9\textwidth]{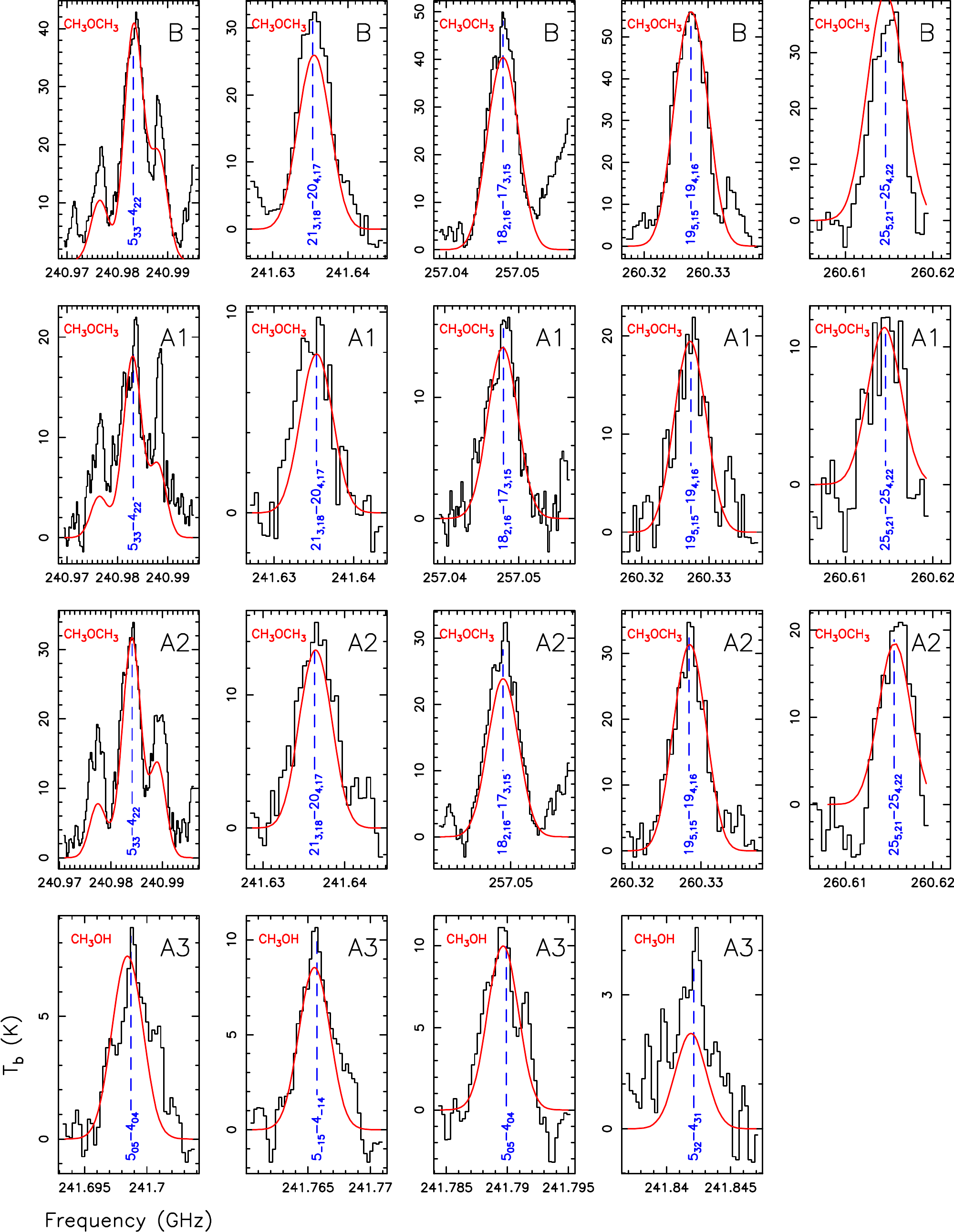}
\caption{In each panel, the observed spectrum and the \textsc{MADCUBA} LTE best fit are shown in black and red, respectively. The spectra are shown as brightness temperature vs. rest frequency. In the upper-left corner and at the bottom of each panel, the molecular species and the corresponding line employed in the fit are indicated (see Table~\ref{used_lines}). Each raw refers to one of the four mm sources:\ B, A1, A2 and A3, from top to bottom, respectively. }
\label{MAD-fit}
\end{figure*}

 Figure~\ref{cont-spe} shows the spectra of the \ CH$_3$CN \ and \ CH$_3$OH \ lines prominent in our ALMA frequency setup, extracted at the position of the 1.2-mm peak for each of the four continuum sources. In agreement with previous findings, the molecular emission is significantly stronger towards B (by a factor of 2.5--5). While in sources~A1~and~A2 the intensity of the \ CH$_3$CN \ and \ CH$_3$OH \ lines is comparable, in A3 only weak \ CH$_3$OH \ emission is detected. Table~\ref{used_lines} lists all the molecular lines analyzed in this paper. The intense lines of \ CH$_3$CN \ and \ CH$_3$OH \ are mostly employed to trace the gas kinematics inside B.
While the optical depth of these lines close to the peak  approaches 1 in some cases, the line wings are always optically thin and allow us to map the gas kinematics around the high-mass YSO in B with good S/N. 
Using the SLIM (Spectral Line Identification and Modeling) tool of \textsc{MADCUBA}\footnote{Madrid Data Cube Analysis on Image (MADCUBA) is a software developed in the Center of Astrobiology (Madrid, INTA-CSIC) to visualize and analyze astronomical (single) spectra and data cubes (Mart\'in et al, in prep.; \citealt{Riv16}); website: http://cab.inta-csic.es/madcuba/Portada.html.}, we surveyed the spectra of the four mm~sources to search for molecular species with a relatively large number of unblended and optically thin lines suitable for deriving the gas physical conditions. We selected the transitions of the \ CH$_3$OCH$_3$ \ and \ CH$_3$OH\footnote{CH$_3$OH \ transitions are used to derive the physical conditions in A3 only, since in this source the emission of \ CH$_3$OCH$_3$ \ is not detected.} \ molecules reported in boldface characters in Table~\ref{used_lines}, which have optical depths of less (or much less) than\ 0.5, and cover a sufficiently wide range in excitation energy to guarantee a good estimate of the excitation temperature. 
To derive the physical parameters we used the tool AUTOFIT of \textsc{MADCUBA}, which compares the observed spectra with the LTE synthetic spectra, taking into account all transitions and the line opacities. Leaving four parameters free to vary, that is, column density ($N_{\rm col}$), excitation temperature ($T_{\rm ex}$), Local Standard of Rest (LSR) velocity (\Vlsr), and line width (FWHM), AUTOFIT  provides the best nonlinear least-squared fit using the Levenberg-Marquardt algorithm. 
Figure~\ref{MAD-fit} shows that the emission profiles of the unblended CH$_3$OCH$_3$ \ and \ CH$_3$OH \ transitions are reasonably well fitted with \textsc{MADCUBA}, determining the values of column density, temperature, velocity and line width listed in Table~\ref{cont-sour}. The masses in molecular gas of the mm sources are derived from the mm fluxes (integrating inside the contours at 10\% of the peak marked in Fig.~\ref{cont}) and the fitted gas temperatures, assuming a dust opacity of \ 1~cm$^2$~g$^{-1}$ at 1.4~mm \citep{Oss94} and a gas-to-dust mass ratio of 100. 
We used the mm continuum peak flux and the fitted gas temperature to estimate the optical depth of the dust emission inside the mm sources,
ranging from \ 0.25 for source~B to \ 0.45 for source~A3. For each mm~source, the mass value reported in Table~\ref{cont-sour} was corrected using the corresponding dust opacity. 
The flux error is evaluated by considering the contributions of the rms noise of the dust continuum map close to the intense sources, $\approx$0.4~m\Jyb, and the ALMA flux calibration uncertainty of \ 5\%, as recently estimated from the analysis of calibrators in bands 3 and 6 by \citet[][see also references therein]{Bon18}. The uncertainty in the mass of the mm~sources is derived by taking into account both the flux and gas temperature uncertainties.
 

\subsection{The massive YSO Bm}
\label{jet-dis}
%

\begin{figure*}
\includegraphics[width=0.58\textwidth]{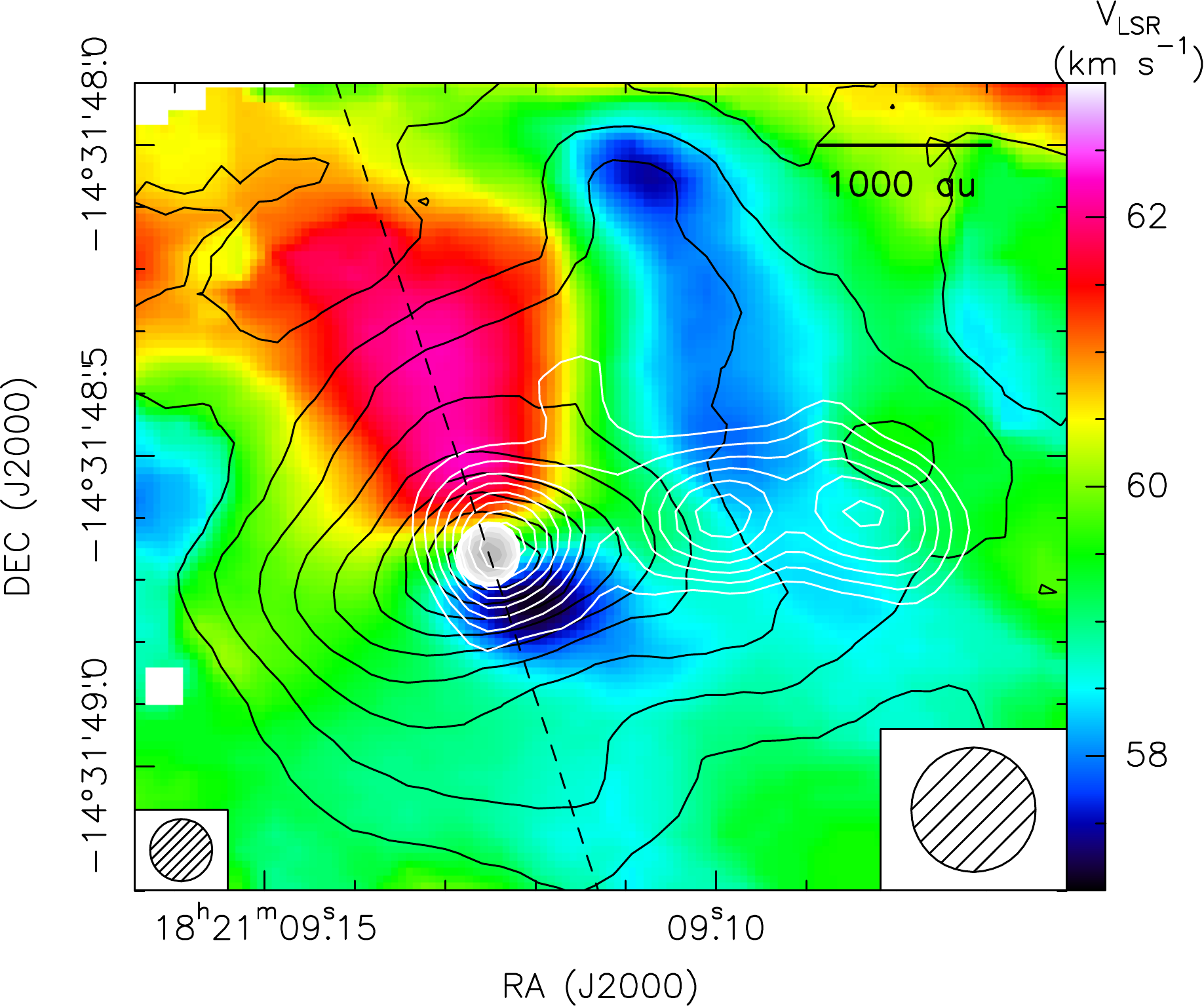}
\includegraphics[width=0.42\textwidth]{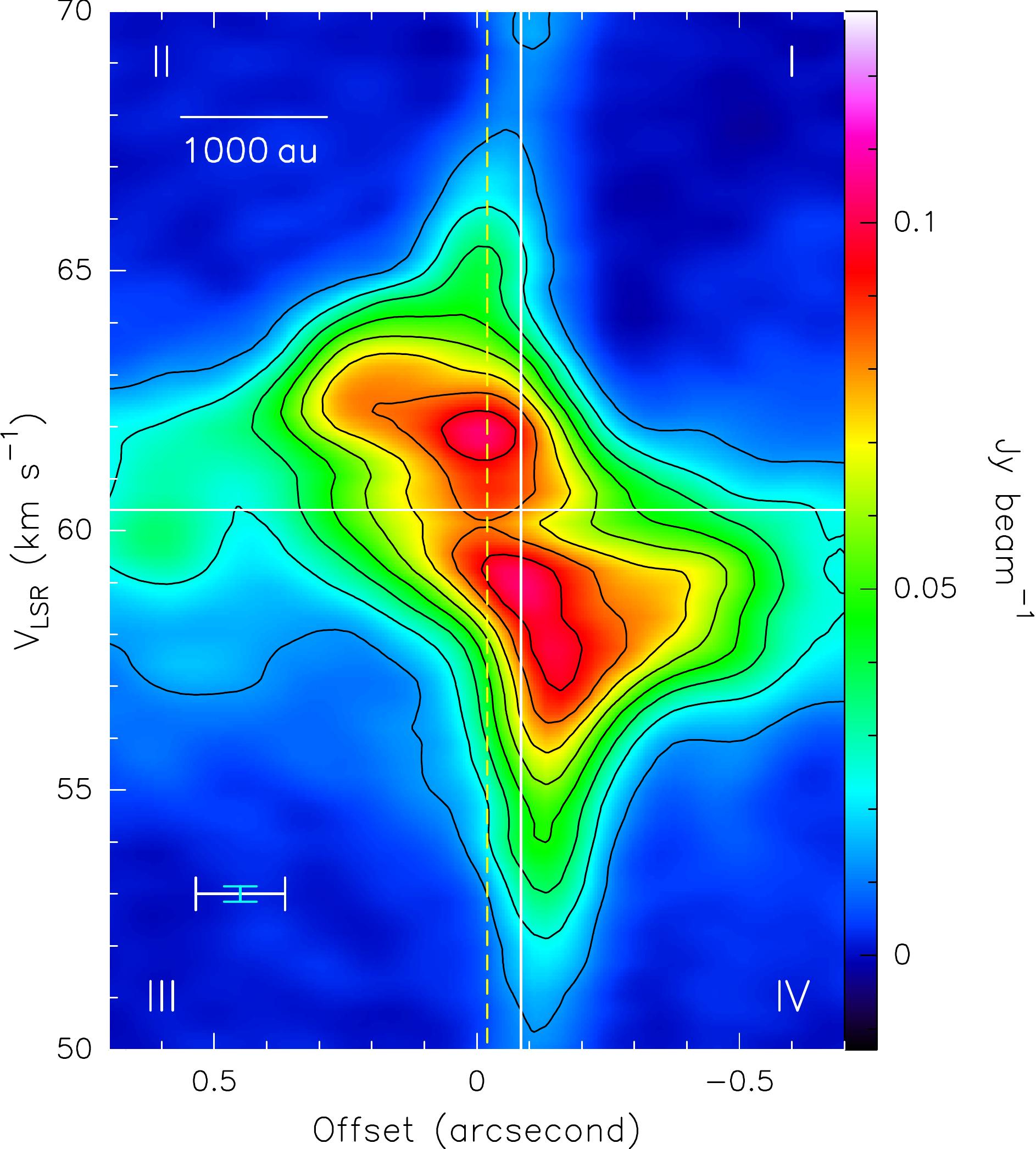}
\caption{Gas kinematics towards the high-mass YSO Bm. \ {\em Left~panel:}~Black contours and the color map show, respectively, the velocity-integrated intensity and the intensity-averaged velocity of the \ CH$_3$OH 5$_{0,5}$--4$_{0,4}$ line (E$_u$ = 35~K). Plotted contours are from 10 to 90\%  of \ 1.1~\Jyb~\kms\ in steps of 10\% and the scale on the right gives the color-velocity conversion. The gray-scale filled and white contours show the JVLA 22-GHz and 13-GHz continuum emission, respectively. Contours are from 50 to 90\% of \ 0.28~m\Jyb\ in  steps of
10\% at 22~GHz, and from 20 to 90\%  of \ 0.1~m\Jyb\ in steps of 10\% at 13~GHz. The JVLA beams at 22~and~13~GHz are reported in the bottom-left and bottom-right corners, respectively. \ {\em Right~panel:}~P-V plot of the CH$_3$OH 5$_{0,5}$--4$_{0,4}$ line. The cut (at PA = 18\degr) along which positions are evaluated is indicated with the dashed black line in the left panel. To produce the P-V plot, we averaged the emission inside a strip parallel to the cut and \ 0\pas4 in width to include only the disk--envelope around the YSO. The intensity scale is shown on the right. The spatial and velocity resolutions are indicated with horizontal white and vertical cyan error bars, respectively, in the bottom-left corner. The horizontal and vertical white continuous lines mark the YSO \Vlsr\ and positional offset, respectively, estimated with a Keplerian fit (see Sect.~\ref{disk-pro}). The yellow dashed vertical line indicates the positional offset of the peak of the 1.2-mm continuum emission of source~B. The four quadrants are labeled.
}
\label{FM_CH3OH}
\end{figure*}

\begin{figure*}
\includegraphics[width=0.58\textwidth]{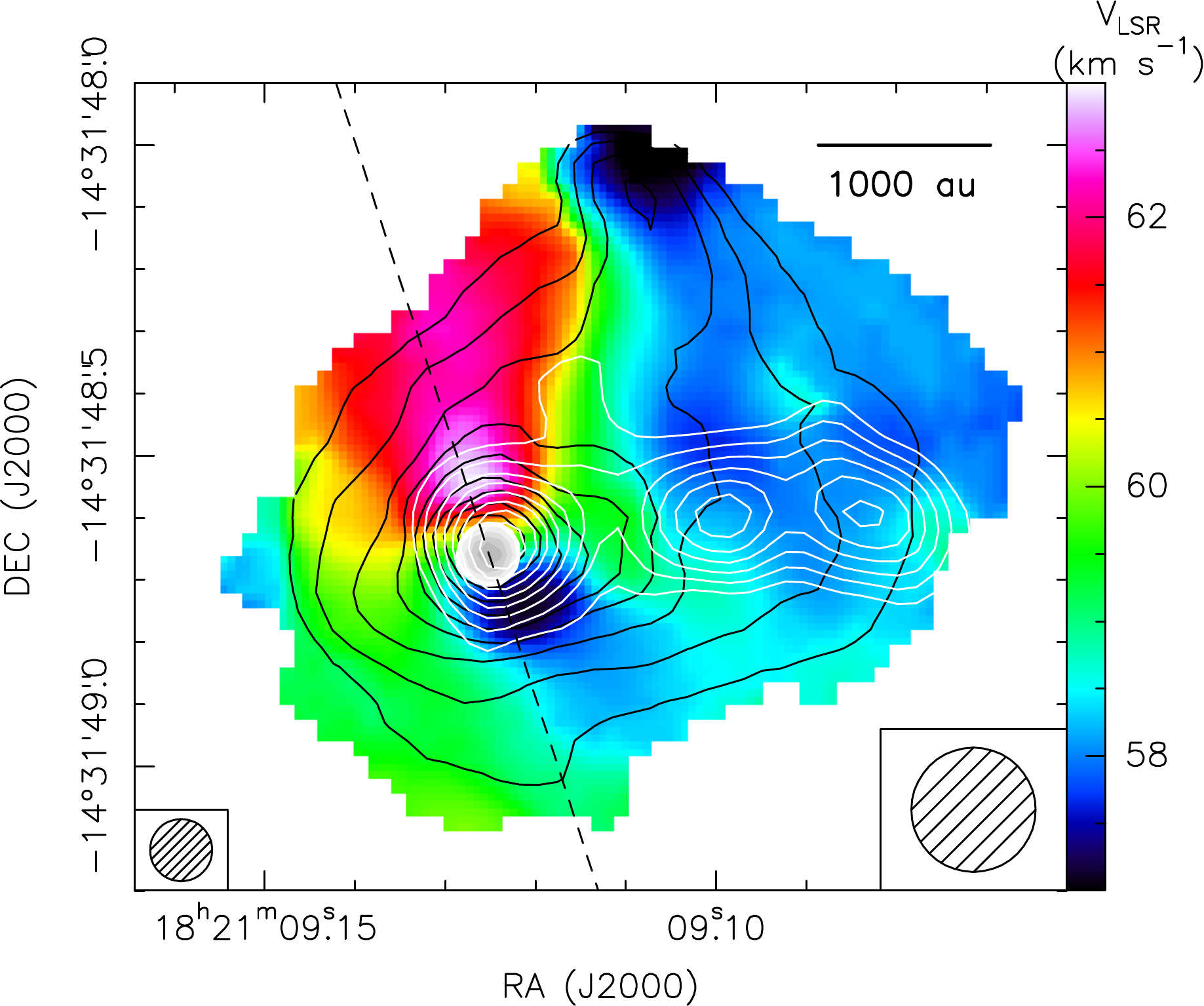}
\includegraphics[width=0.42\textwidth]{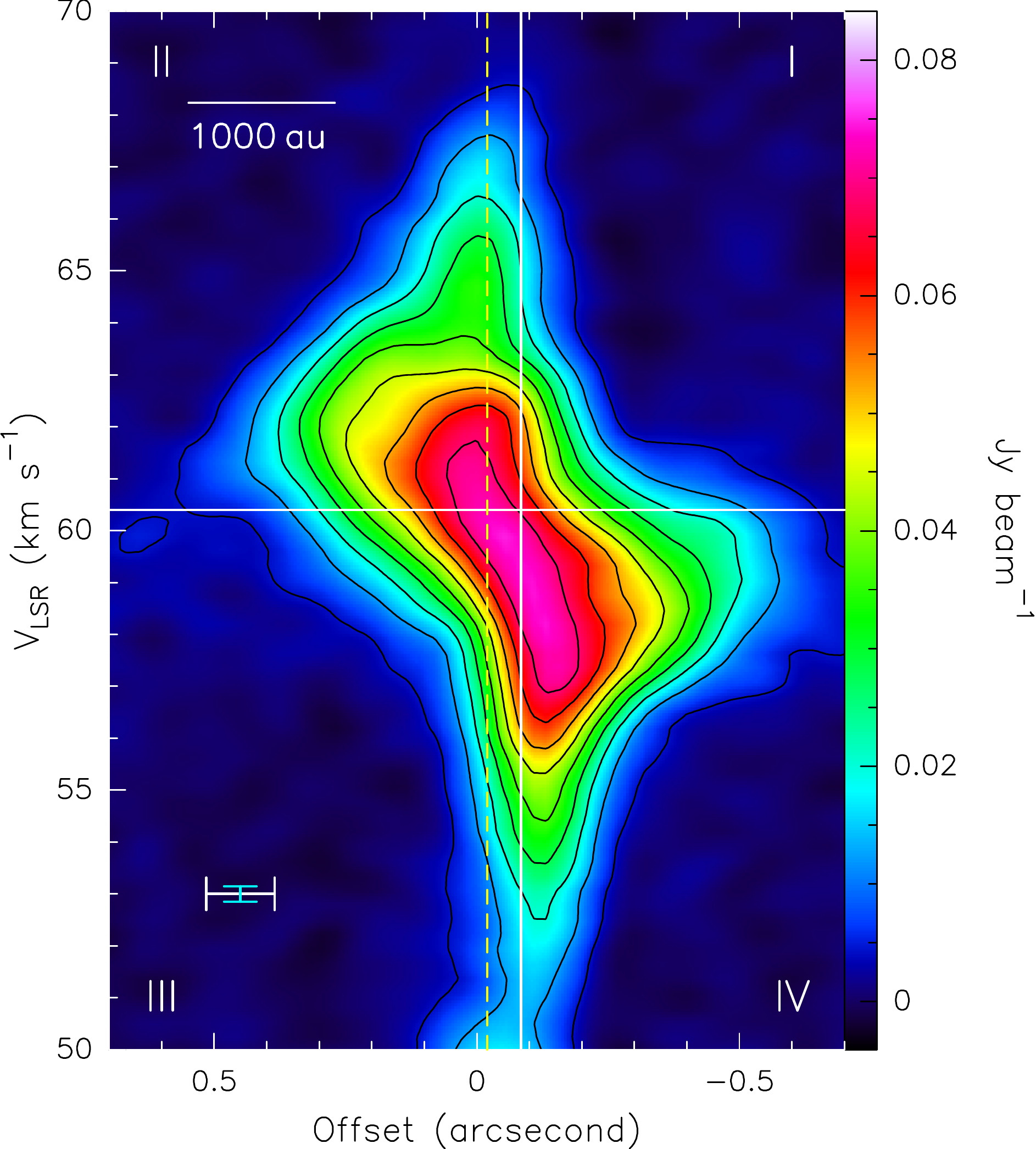}
\caption{As in Fig.~\ref{FM_CH3OH} but for the \ CH$_3$CN J$_K$~=~14$_2$--13$_2$ line (E$_u$ = 121~K).  \ {\em Left~panel:}~Black contours show  the velocity-integrated intensity with levels ranging from  10 to 90\% of \ 0.9~\Jyb~\kms \ in steps of 10\%.}
\label{FM_CH3CN}
\end{figure*}

\begin{figure*}
\includegraphics[width=0.58\textwidth]{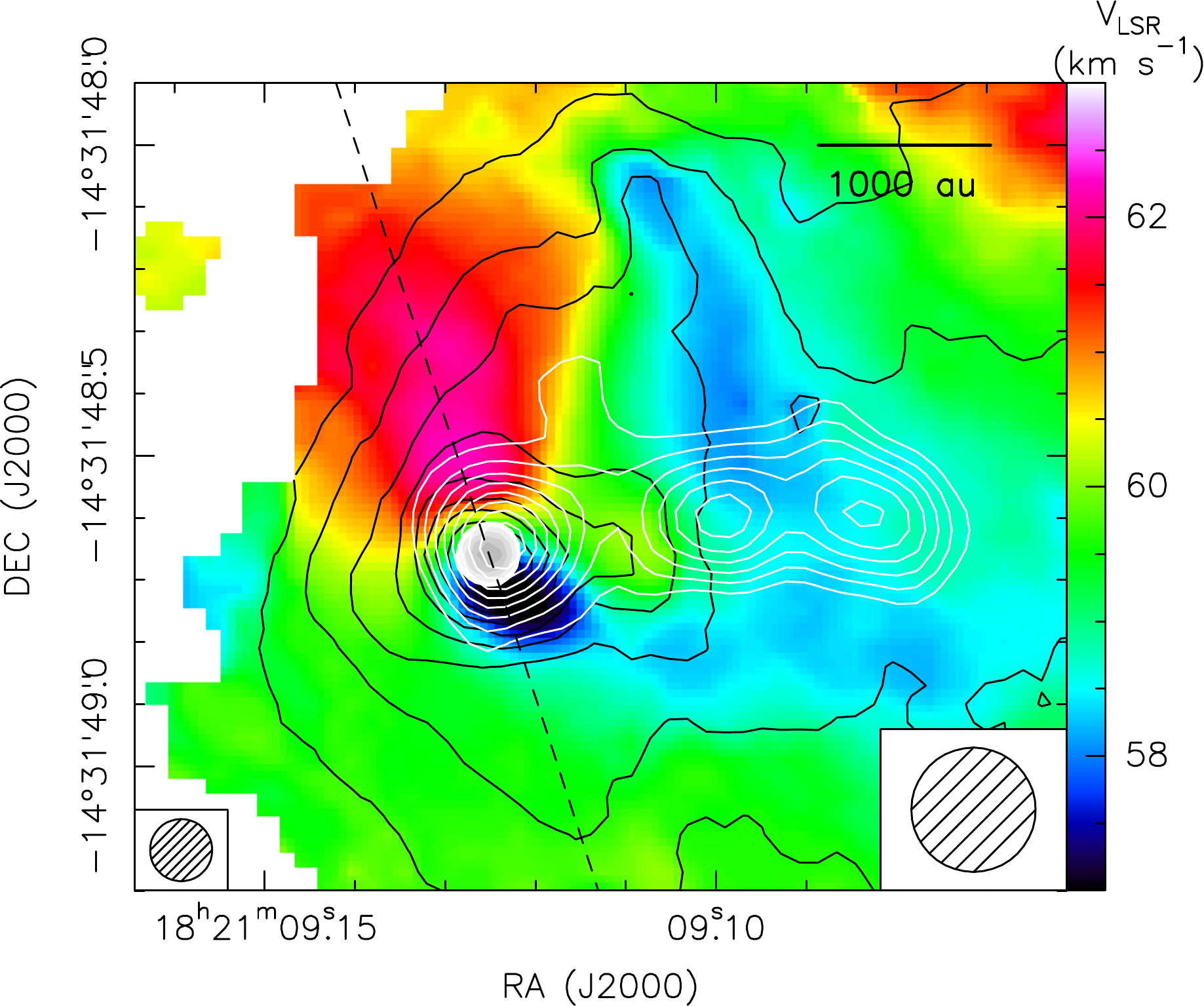}
\includegraphics[width=0.42\textwidth]{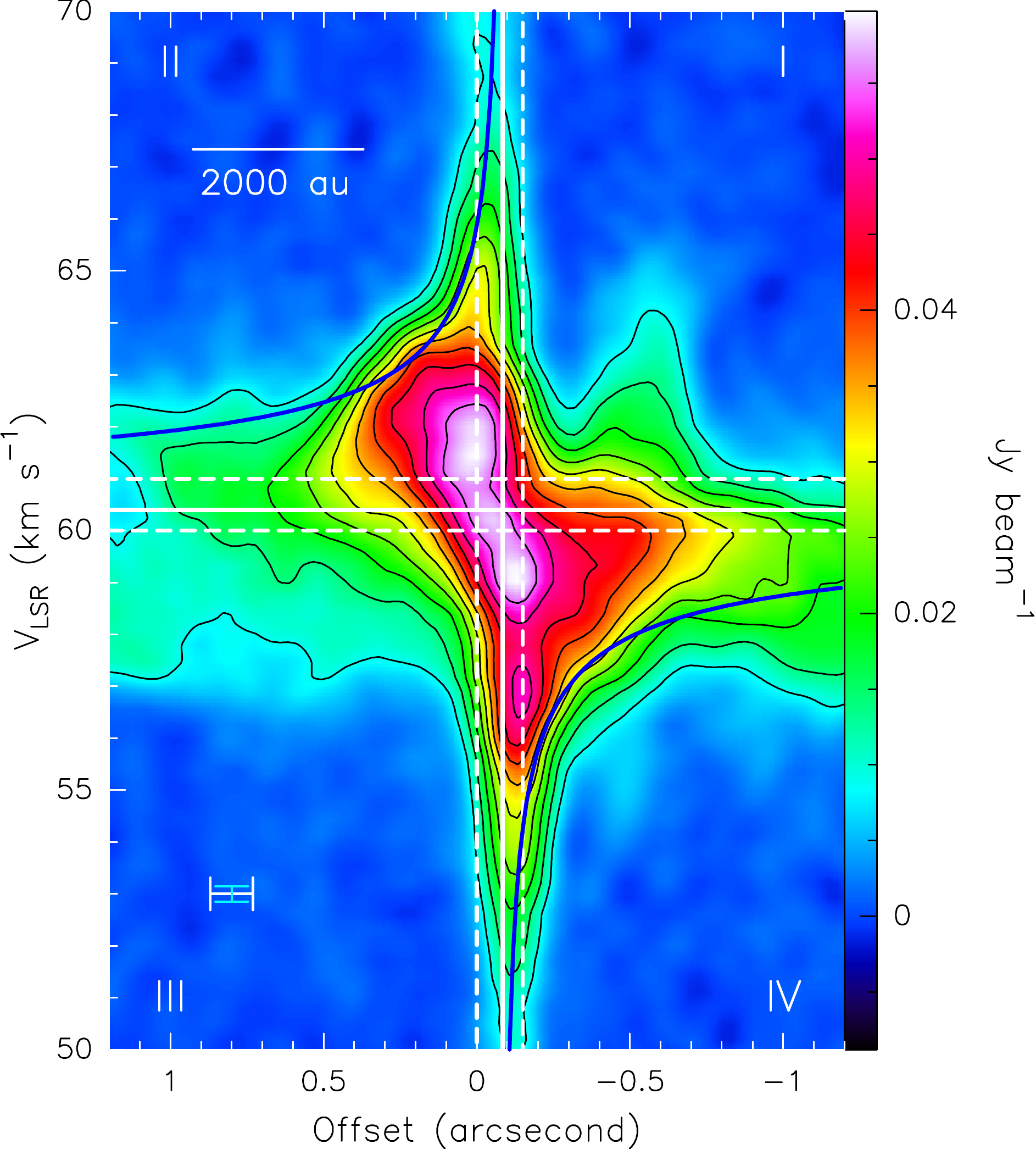}
\caption{As in Fig.~\ref{FM_CH3OH} but for the \ C$^{34}$S J~=~5--4 line (E$_u$ = 28~K).  \ {\em Left~panel:}~Black contours show the velocity-integrated intensity with levels ranging from  10 to 90\% of \ 0.8~\Jyb~\kms \ in steps of 10\%. \ {\em Right~panel:}~Horizontal and vertical white continuous lines, and the blue curves mark, respectively, the \Vlsr\ and positional offset of the YSO, and the Keplerian profile around a YSO of \ 10~\ms. The two horizontal and vertical, white dashed lines indicate the maximum interval of variation for the YSO \Vlsr\ and  positional offset, respectively, estimated by eye whilst looking at the symmetrical patterns of the P-V plot.  We note that the plotted offset range is larger than for the P-V plots of Figs.~\ref{FM_CH3OH}~and~\ref{FM_CH3CN}.}
\label{FM_C34S}
\end{figure*}

We now focus on the gas kinematics around the massive YSO inside the mm source~B. In the following we refer to this YSO as Bm. This is probably the most massive object in source~B. A well-defined, SW-NE  \Vlsr\ gradient is detected at the position of Bm in all the observed high-density gas tracers. Figures~\ref{FM_CH3OH},~\ref{FM_CH3CN},~and~\ref{FM_C34S} present results from low-E$_u$ lines of CH$_3$OH, CH$_3$CN, and C$^{34}$S, respectively. We note that the JVLA 22-GHz continuum, pinpointing the YSO, falls exactly between the SW blue-shifted and the NE red-shifted line emission, and that the direction of the \Vlsr\ gradient forms a large ($\approx$70\degr) angle with the radio jet traced by the extended JVLA 13-GHz continuum. These findings suggest that the \Vlsr\ gradient could be due to rotation of the gas in an envelope and/or disk surrounding the YSO.

The P-V plots produced along the axis (at PA = 18\degr) of the \Vlsr\ gradient confirm that we are observing envelope--disk rotation. The plots have a butterfly-like shape (mostly clear in Fig.~\ref{FM_C34S} for the C$^{34}$S line), with well-defined spurs at high absolute velocities and small offsets in the second and fourth quadrants. These spurs correspond to gas whose  line-of-sight velocity  increases with decreasing radius and could be consistent with Keplerian rotation. In Sect.~\ref{disk-pro}, we fit the velocity profile and estimate the disk radius and mass, and the YSO mass. Here, we constrain the YSO \Vlsr\ and position by inspecting the P-V plots. The positional offset (along the axis at PA = 18\degr) of the YSO is delimited by the offsets of the high-velocity spurs in the second and fourth quadrants of the P-V plots, falling inside the interval \ 0\arcsec--0\farcs15 (see Fig.~\ref{FM_C34S}). The spurs at low (line of sight) velocities and large offsets are less clear, because they are contaminated by emission of the ambient gas at the systemic velocity. We make use of the P-V plot of the C$^{34}$S line (see Fig.~\ref{FM_C34S}) to confidently set a range of values for the YSO \Vlsr\ within \ 60--61~\kms.  These constraints for the YSO position and the \Vlsr\ are used in the following analysis.

Looking at Figs.~\ref{FM_CH3OH},~\ref{FM_CH3CN},~and~\ref{FM_C34S} (left panels), in all the three molecular tracers the blue-shifted emission extends over an area significantly smaller than that of the red-shifted emission.
Moreover, while the red-shifted emission is distributed approximately symmetrically about the major axis of the envelope--disk, the barycenter of the blue-shifted gas is offset to W. These asymmetries in the velocity distribution of the gas around Bm are further discussed in 
Sects.~\ref{disk-jet}~and~\ref{pm-pre}.

\subsection{The surroundings of Bm}
\label{jet-mol}
\begin{figure*}
\centering
\includegraphics[width=\textwidth]{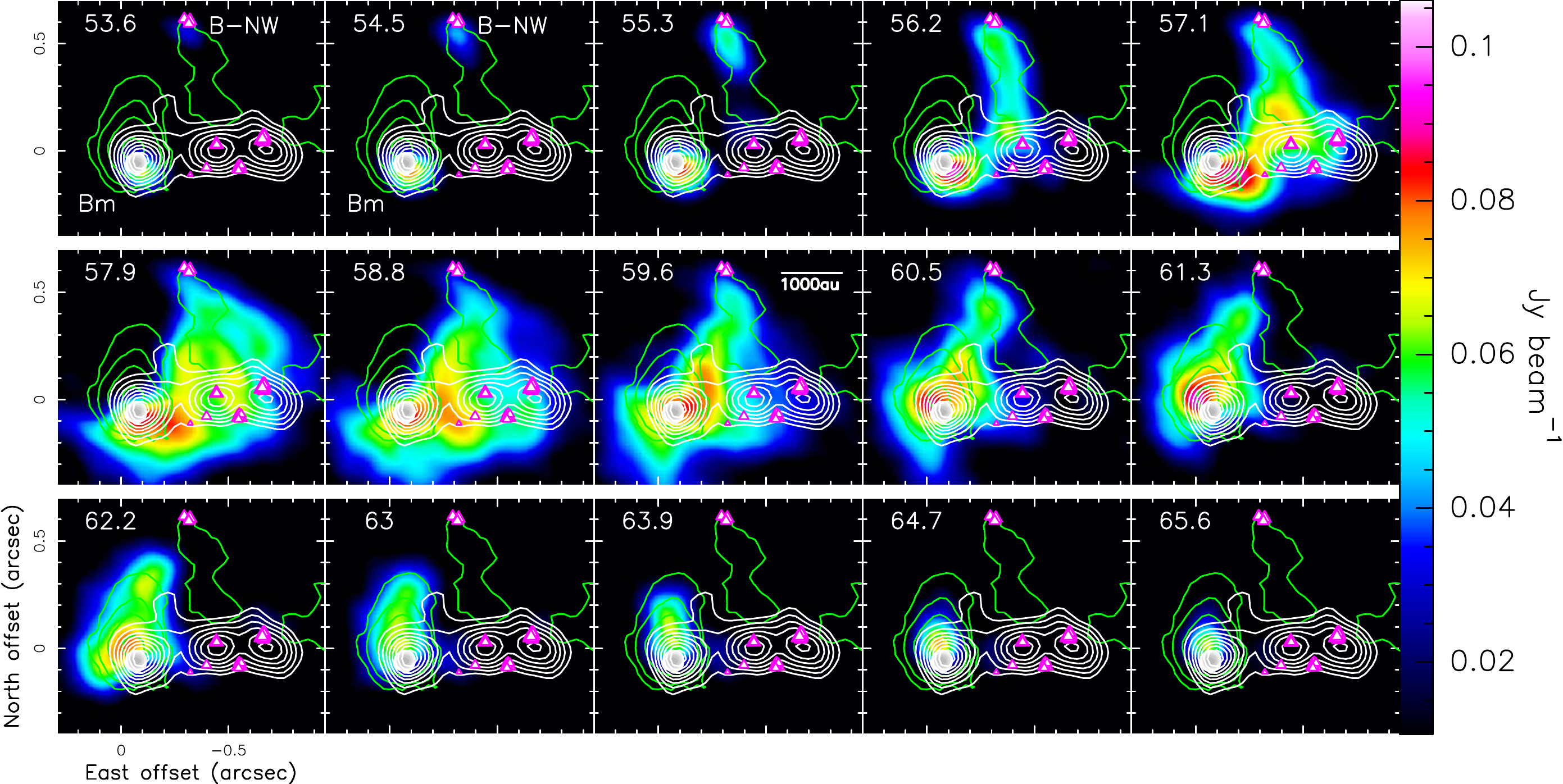}
\caption{Each panel presents a color map of the intensity of the CH$_3$CN J$_K$ =  14$_2$--13$_2$ line (E$_u$ = 121~K) at a different \Vlsr\ (\kms), indicated in the upper-left corner. The color-intensity scale is shown on the right. The grayscale-filled and white contours show the JVLA 22-GHz and 13-GHz continuum, respectively, plotting the same levels as in Fig.~\ref{FM_CH3OH}. The green contours reproduce the ALMA 1.2-mm continuum, showing levels at 10\%, 20\%, 40\%, and 80\% of the peak emission of 0.032~\Jyb. The magenta-edged white triangles mark the positions of the 22-GHz water masers derived through VLBI observations by \citet{San10a}. In the upper first and second panels, the high-mass YSO Bm, and the compact molecular source B-NW  are labeled.}
\label{W-NW_K2}
\end{figure*}

\begin{figure*}
\centering
\includegraphics[width=\textwidth]{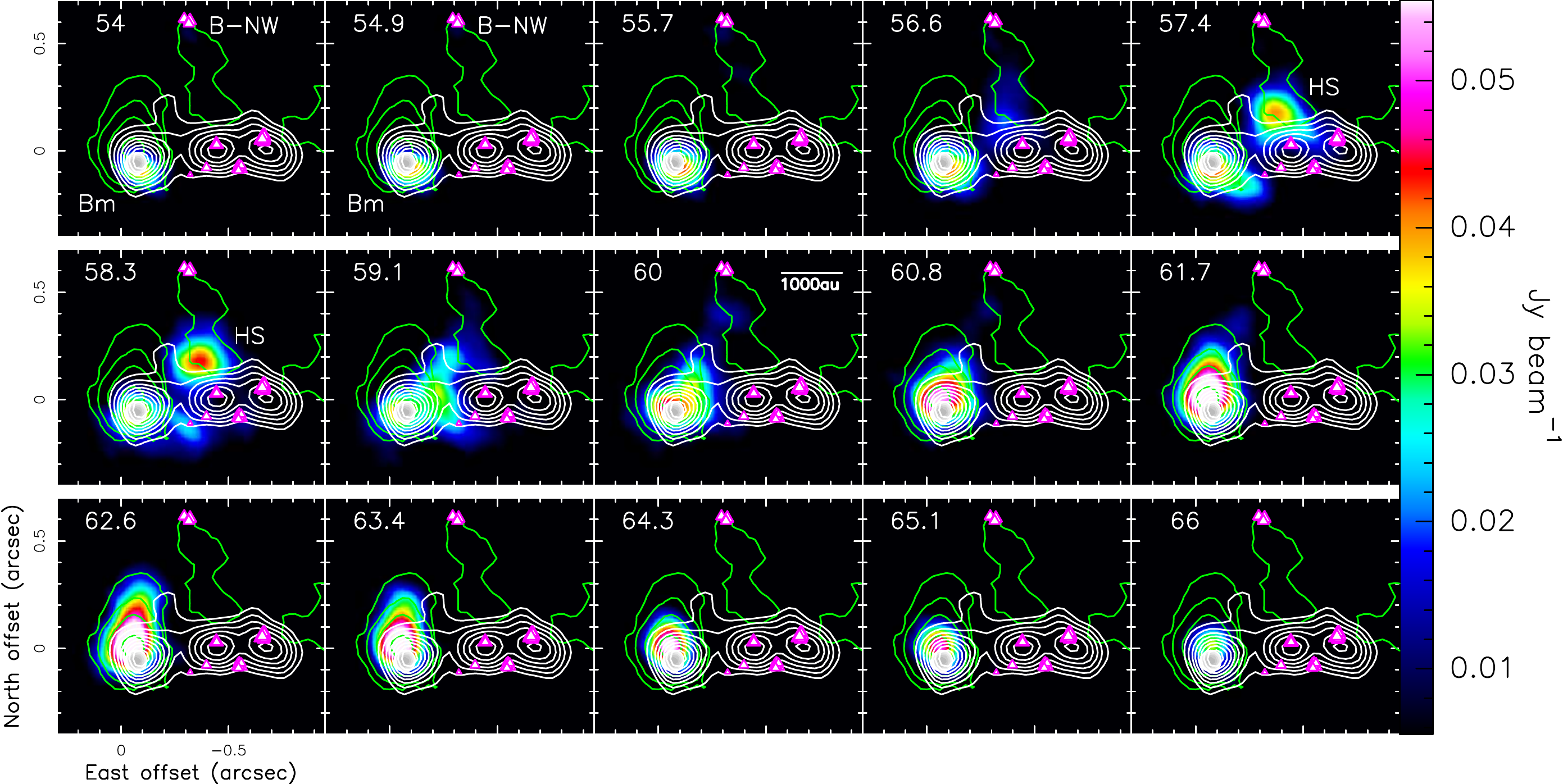}
\caption{As in Fig.~\ref{W-NW_K2} but for the \ CH$_3$CN J$_K$ =  14$_7$--13$_7$ line (E$_u$ = 442~K).  In the upper-last and middle-first panels, the emission spot~HS, adjacent to the radio jet, is labeled.}
\label{W-NW_K7}
\end{figure*}

We now consider a larger region encompassing the jet emerging from Bm. Figures~\ref{W-NW_K2}~and~\ref{W-NW_K7} show the channel maps of two intense CH$_3$CN lines with quite different excitations, the \ J$_K$ =  14$_2$--13$_2$ with \ $E_u/k$ = 121~K \ and the \ J$_K$ =  14$_7$--13$_7$ with \ $E_u/k$ = 442~K. These two lines are representative of the velocity distribution of the large majority of the low-~and~high-excitation molecular transitions detected towards Bm. At high velocities (channels with \Vlsr \ $\le$~56~\kms \ or \ $\ge$~63~\kms) both low- and high-excitation molecular lines trace the compact disk close to the high-mass YSO. At \Vlsr $\approx$55~\kms, a compact source is detected only in low-excitation lines at \ $\approx$0\farcs6 towards N-NW of the mm~peak of source~B, in proximity to the weak continuum emission and the cluster of water masers located at the largest distance from the radio jet. In the following, we refer to this compact source as \ B-NW. At central velocities (56~\kms~$\le$ \Vlsr $\le$~63~\kms), low-excitation lines show extended emission emerging from (only) the western side of the jet at blue-shifted velocity, and from an arc-like/linear bridge at blue-/red-shifted velocities connecting the massive YSO Bm with the compact source~B-NW. High-excitation lines have a quasi-compact structure also at central velocities, with some notable features:\ 1)~at \Vlsr$\approx$59~\kms, a slightly resolved funnel-like structure is traced on top of the western lobe of the jet near Bm, and \ 2)~at \Vlsr $\approx$57-58~\kms,  an emission spot is observed adjacent to the radio jet towards N. This latter feature is referred to as HS in the following.

The radio jet from Bm is one of the few known cases of nonthermal jets from high-mass YSOs. In particular, the eastern lobe of the radio jet is well detected with the JVLA at 6~GHz, and is very weak and/or undetected at frequencies  $\ge$10~GHz \citep[][see their Figs.~1~and~2]{Mos13b}. Our ALMA observations confirm the presence of a strong asymmetry between the eastern and western lobes of the jet, with the western lobe being much more intense in both continuum and molecular lines. Considering also the distribution of the dust emission (see Fig.~\ref{cont}), predominant towards W-NW of Bm, the simplest explanation for all these facts is an E-W density gradient at the position of Bm. Integrating the ALMA 1.2-mm continuum inside the area of each jet lobe, we estimate that the average gas density in the western lobe is about one order of magnitude larger than in the eastern lobe.



\section{The Keplerian disk around the high-mass YSO Bm}
\label{disk-pro}
\begin{figure*}
\includegraphics[width=0.52\textwidth]{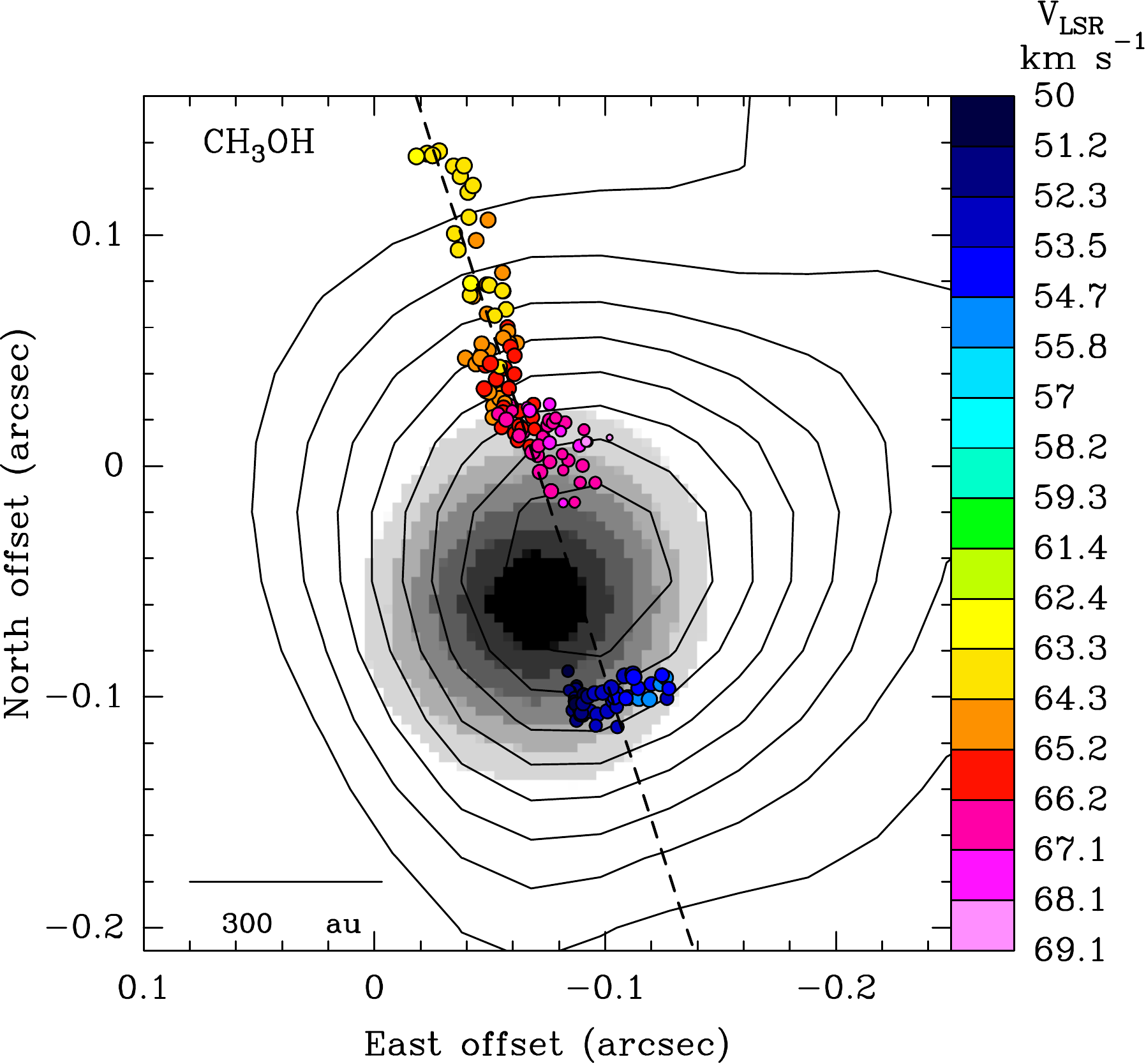} 
\includegraphics[width=0.495\textwidth]{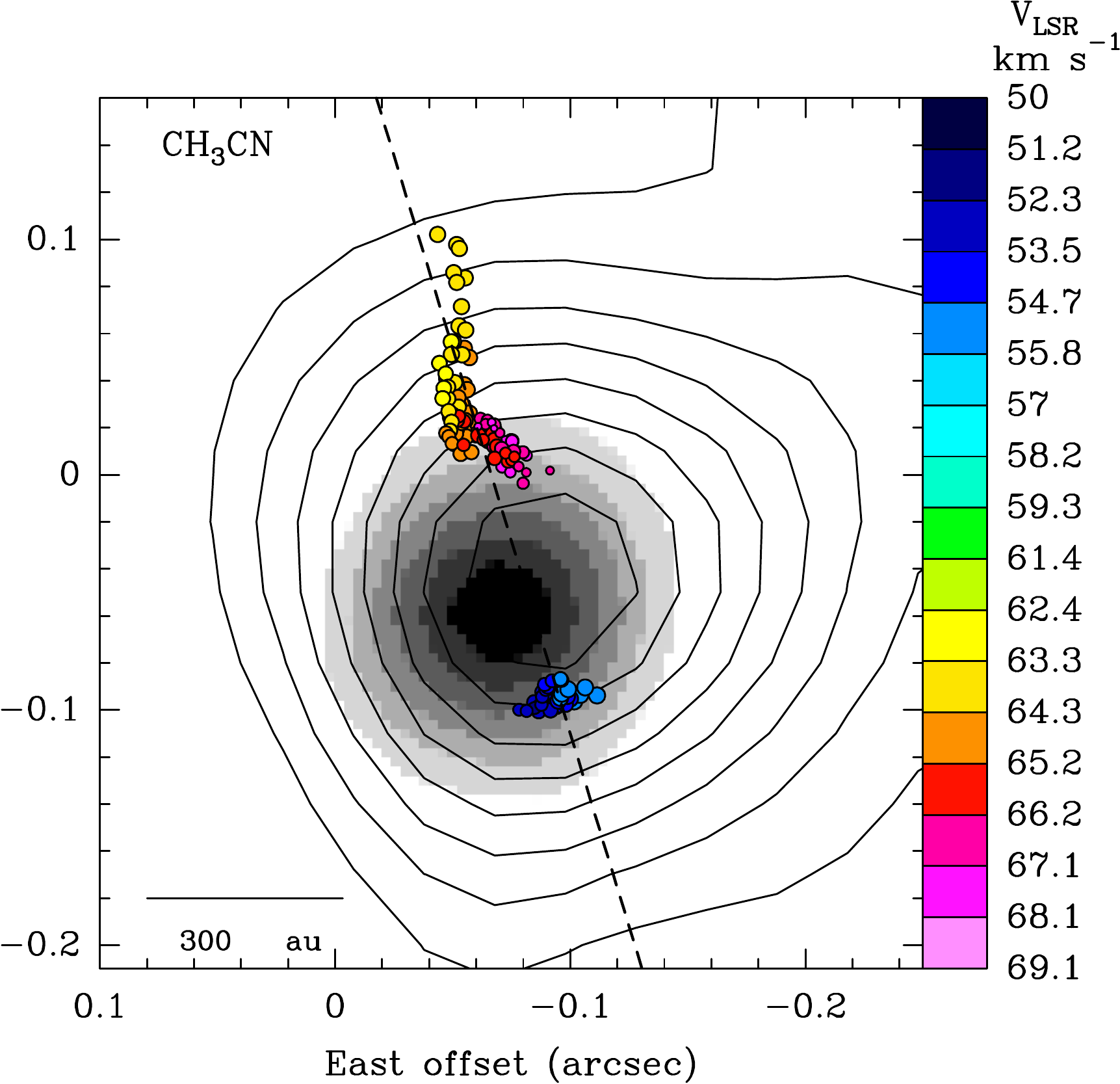}
\caption{The disk around the high-mass YSO Bm. Colored dots indicate the peak positions of the most blue-~and~red-shifted velocity channels for the emission of the nine CH$_3$OH ({\em left~panel}) and seven CH$_3$CN ({\em right~panel}) lines listed in Table~\ref{used_lines}. Colors represent \Vlsr\ as coded on the right of each panel. The dashed black lines, at PA = 18\degr and 17\degr \ for CH$_3$OH and CH$_3$CN, respectively, show the linear fits to the spatial distribution of the channel peaks. The grayscale-filled and black contours represent the JVLA 22-GHz and 13-GHz continuum, respectively, with the same levels as in Fig.~\ref{FM_CH3OH}.}
\label{peak_disk}
\end{figure*}

In this section, we derive the properties of the YSO Bm and its disk.
The angular resolution of the P-V plots shown in Figs.~\ref{FM_CH3OH},~\ref{FM_CH3CN},~and~\ref{FM_C34S} varies in the range \ 0\farcs13--0\farcs17, reflecting the FWHM of the observing beam at the frequency of the molecular line employed to produce the P-V plot. The emission at high blue- ($\le$56~\kms) and red-shifted ($\ge$63~\kms) velocities is approximatively compact (see Figs.~\ref{W-NW_K2}~and~\ref{W-NW_K7}), which allows us to fit a 2D  Gaussian profile and  determine the peak position at each velocity channel. The positional accuracy is equal to \ \ $ \frac{\rm FWHM}{2.0} \ \frac{\sigma}{I} $ \citep[see, e.g.,][]{Rei88}, where \ $I$ \ and \ $\sigma$ \ are the peak intensity and the rms noise, respectively, of a given channel. With typical values of the ratio \ $ \frac{I}{\sigma} \approx 30 $, the error on the position of the fitted channel peaks is as small as a few mas. For channels where the emission is sufficiently compact, this method increases  the (relative) positional accuracy by a large factor, while retaining the essential information of the P-V plots.

In Fig.~\ref{peak_disk} we show that the spatial distribution of the high-velocity emission of CH$_3$OH and CH$_3$CN is elongated along the same \ PA = 18$\pm$3\degr, which we take as the direction of the major axis of the molecular disk around the high-mass YSO Bm. As already noted in Sect.~\ref{jet-dis} for the SW-NE \Vlsr\ gradient, the red-shifted side of the disk is significantly longer than the blue-shifted side. Along the red-shifted side, $\approx$0\farcs15 \ or \ $\approx$500~au in extent, the gas \Vlsr\ increases monotonically approaching the YSO, ranging from \ $\approx$62~\kms\ up to \ $\approx$69~\kms. On the contrary, the blue-shifted side of the disk is traced only at the lowest velocities, 50--55~\kms; the mildly blue-shifted channels, in the range \ 55--60~\kms, show more extended structure, which tends to be elongated to the W of Bm along the radio jet (see Fig.~\ref{W-NW_K2}). 

\begin{figure*}
\centering
\includegraphics[width=\textwidth]{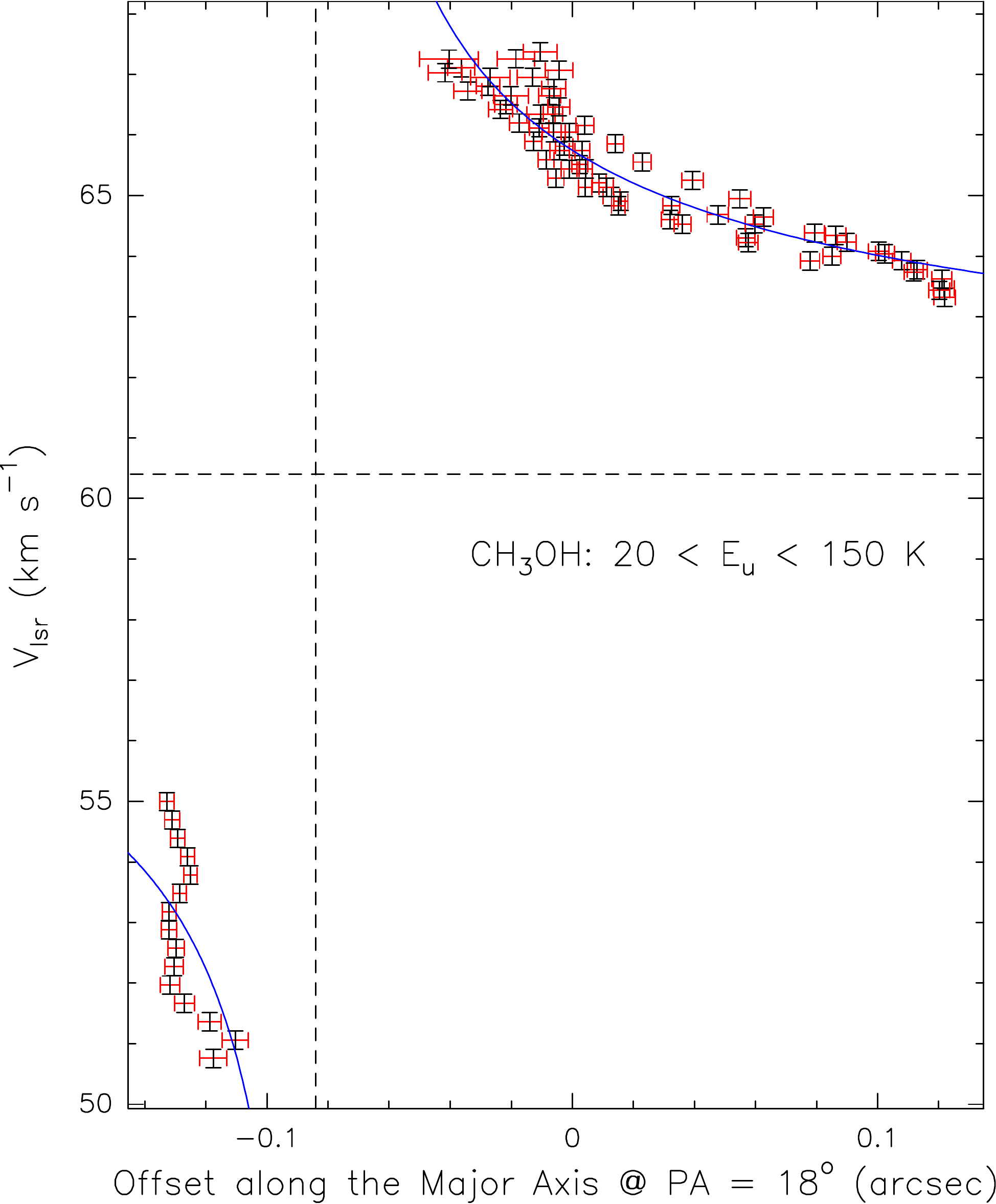}
\caption{Black and red error bars give major-axis projected positions and \Vlsr\ (together with the corresponding errors), respectively, for the highest-velocity emission peaks of the six \ CH$_3$OH lines with \ 20~K~$\le E_u \le$~150~K listed in Table~\ref{used_lines}. The blue curve is the best Keplerian fit to the data. The horizontal and vertical dashed lines indicate the fitted YSO \Vlsr\ and position, respectively. }
\label{Chi2-fit}
\end{figure*}

Comparing the two molecular tracers in Fig.~\ref{peak_disk}, the red-shifted side of the disk appears to be better traced by the CH$_3$OH lines, whose spatial distribution is flatter and more extended than that of the CH$_3$CN lines. Table~\ref{used_lines} shows that the upper energy level (E$_u$) of the CH$_3$OH transitions is on average lower than that of the CH$_3$CN lines. Since the gas temperature in the central layers of the disk is expected to be relatively low, it is reasonable that the CH$_3$OH transitions are found to be better tracers of the disk midplane. Accordingly, examining the different CH$_3$OH transitions (see Table~\ref{used_lines}), we find that the ones with \ E$_u$ $\le$~150~K \  present a  significantly
flatter and more elongated spatial distribution. Considering only the CH$_3$OH transitions with \ 20 $\le$ E$_u$ $\le$~150~K, Fig.~\ref{Chi2-fit} plots \Vlsr\ versus positions projected along the disk major-axis. While the radial profile of the red-shifted velocities is well reproduced with a Keplerian curve (as described below), the blue-shifted emission concentrates within a small offset range ($\le$0\farcs04) and cannot be adequately fitted. However, the average position and \Vlsr\ of the blue-shifted emission is useful to constrain the position and the \Vlsr\ of the YSO. 

\begin{figure*}
\includegraphics[width=0.45\textwidth]{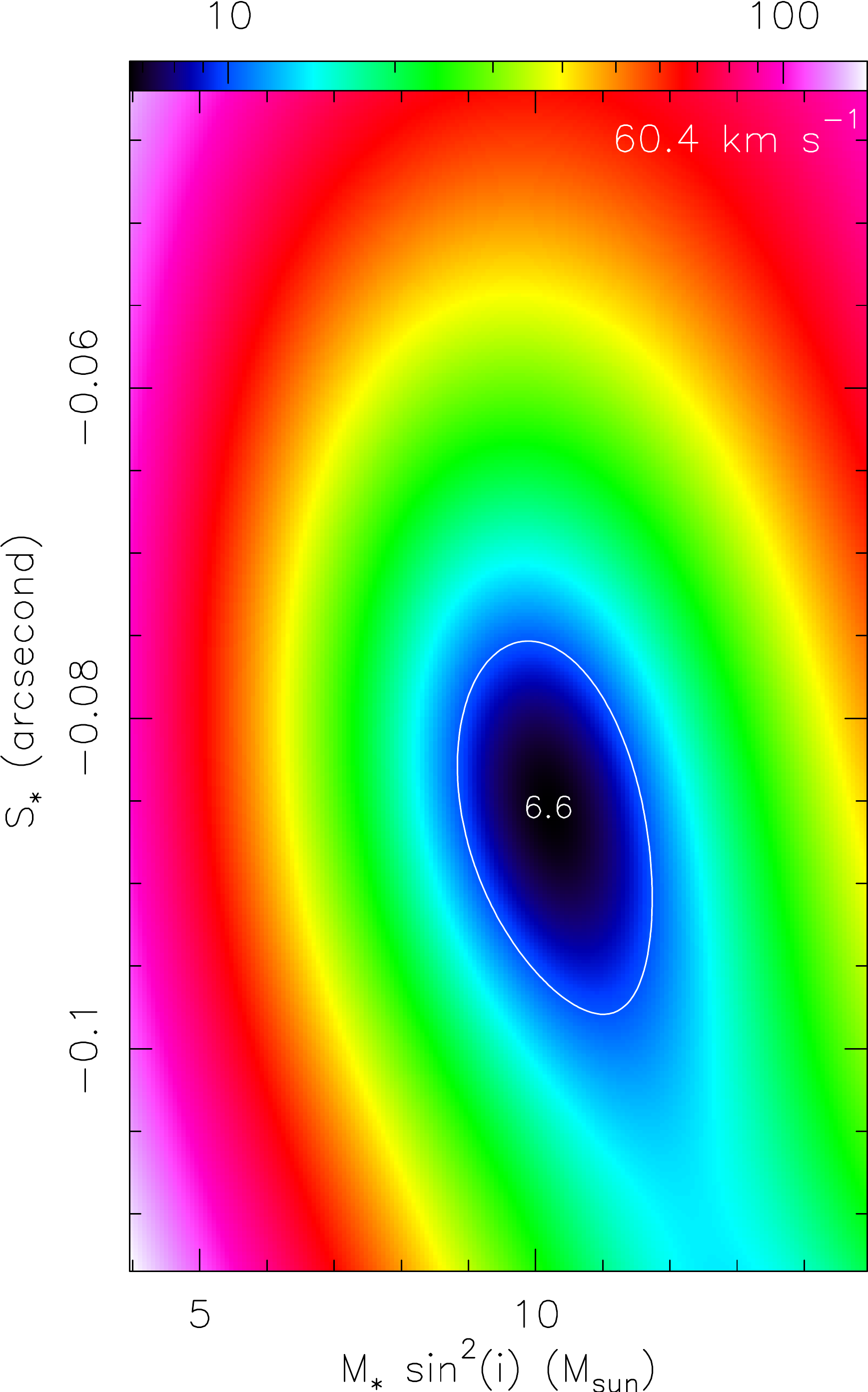} 
\hspace*{0.5cm} \includegraphics[width=0.45\textwidth]{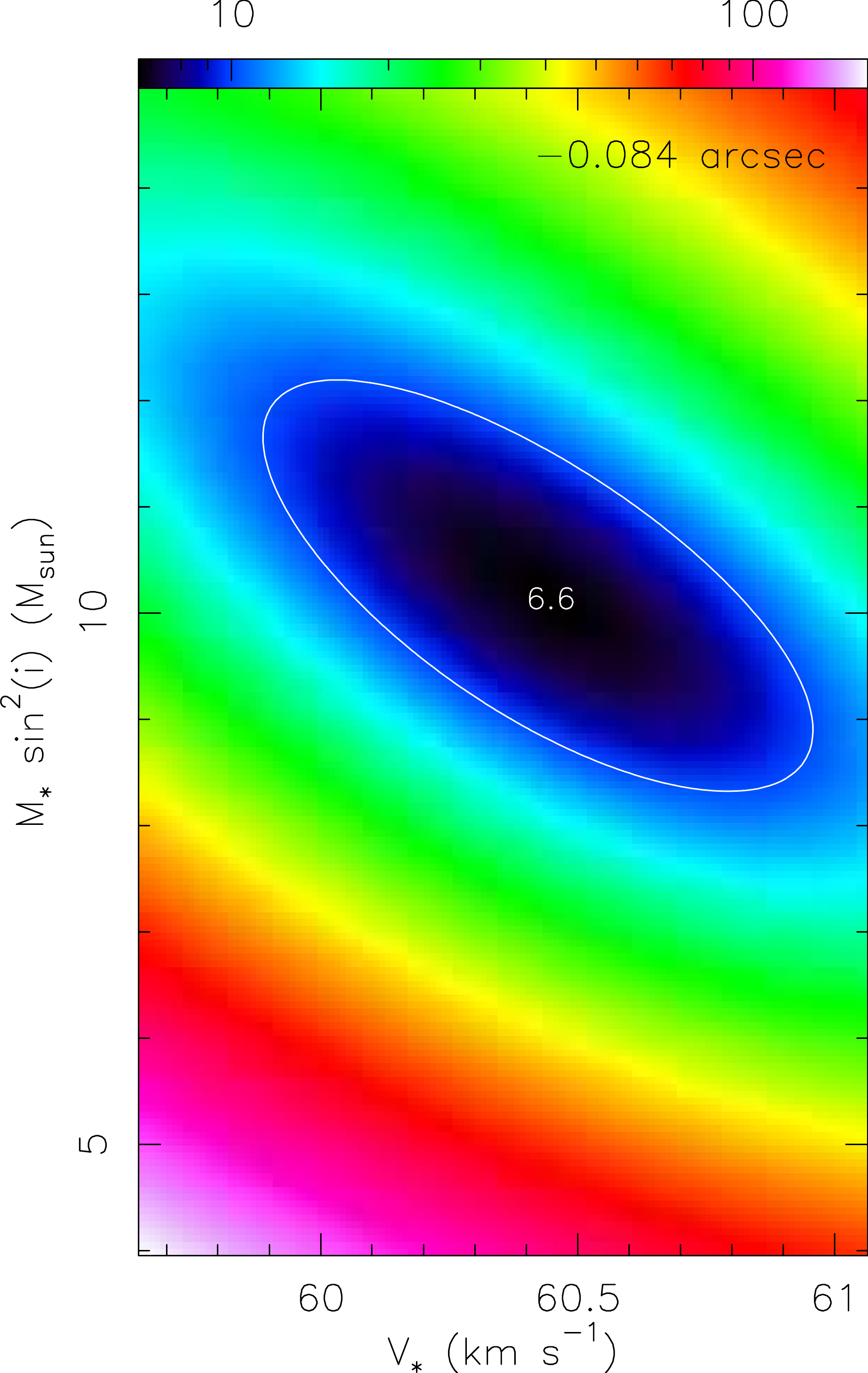}
\caption{Plots of the $\chi^2$ distribution from the Keplerian fit (see Fig.~\ref{Chi2-fit}) as a function of the free parameters \ YSO mass and position ({\em left~panel}), and YSO \Vlsr\ and mass ({\em right~panel}). To produce these plots, the third free parameter of the Keplerian fit, that is, the YSO \Vlsr\ and position in the left and right plot, respectively, is taken equal to the best-fit value. The color scale at the top of each panel gives the value of the $\chi^2$ in a logarithmic scale. In each of the two plots, the position of the minimum \ $\chi^2$, $\chi^2_{min}$ = 6.6, is labeled, and the white contour indicates the level \ $\chi^2_{min}$+3.5 = 10.1, which corresponds to the 1-$\sigma$ confidence level for three free parameters \citep{Lam76}. }
\label{Chi2-plots}
\end{figure*}

The free parameters of the Keplerian fit are the YSO's positional offset, $S_{\star}$, LSR velocity, $V_{\star}$, and mass, $M_{\star} \, \sin^2(i)$. Indicating with \ $i$ \ the inclination of the disk rotation axis with the line of sight, the formulation of the latter parameter takes into account that the disk plane is actually seen at an angle \ $90-i$ \ from the line of sight, and the fitted \Vlsr\ correspond to the actual rotation velocities multiplied by the factor \ $\sin(i)$. We have minimized the following \ $\chi^2$ expression:
\begin{equation}
\chi^2 = \sum_j  \frac{ [ V_j - ( V_{\star} \pm 0.5 \; (M_{\star} \sin^2(i) )^{0.5} \; \;  | S_j-S_{\star} |^{-0.5} \, ) ]^2 } {( \Delta V_{j} )^2} 
,\end{equation}

 where \ $V_j$ \ and \ $S_j$ \ are the channel \Vlsr\ (in \kms) and corresponding peak positions (in arcsecond), the index $j$ runs over all the fitted channels, and the $+$ and $-$ symbols hold for red-~and blue-shifted velocities, respectively. The YSO mass $M_{\star}$ \ is given in \ms. 
 In order to take into account both the uncertainty on the velocity and
that on the position, the global velocity error \ $\Delta V_{j}$ \ was obtained by summing in quadrature two
errors: that on the velocity (taken equal to half of the channel width) and
that obtained by converting the error on the offset into a velocity error
through the function fitted to the data.
 In Sect.~\ref{jet-dis}, based on the symmetrical patterns of the P-V plots, we constrained the ranges for the YSO positional offset and \Vlsr\ to: \  $0\arcsec \le S_{\star} \le 0\farcs15 $,  \ 60.0~\kms\ $\le V_{\star} \le$ \ 61.0~\kms. The mass of the YSO is searched over the range \ 5--15~\ms, consistent with the estimated bolometric luminosity of \ 10$^4$~\ls. Figure~\ref{Chi2-plots} reports the plots of the distribution of the $\chi^2$ \ as a function of the free parameters. The white contour in these two plots draws the 1-$\sigma$ confidence level for the three free parameters, following \citet{Lam76}. These plots show that we do find an absolute minimum of the $\chi^2$, and the determined 1-$\sigma$ intervals for the parameter values are: \ $S_{\star}~=~-0\farcs084~\pm~0\farcs020 $, \ $V_{\star}~=~60.4~\pm~0.5$~\kms, and \ $M_{\star} \, \sin^2(i)~=~10~\pm~2$~\ms.


Following the analysis by \citet[][see in particular their Fig.~5]{Mos13b}, source~B, the HMC, dominates the IR  emission of the region, suggesting that our estimate of 10$^4$~\ls\  for the bolometric luminosity of B is reasonable and not too high. This luminosity corresponds to a single ZAMS star of \ $\approx$13~\ms\ \citep[see][]{Dav11}, which is an upper limit for the mass of Bm. Comparing this upper limit with the YSO mass from the Keplerian fit, the value of the disk inclination angle is constrained within the interval \ $60\degr \le i \le 120\degr$, that is, the plane of the disk is within \ 30\degr \ of the line of sight. However, if a sizeable fraction of the bolometric luminosity is due to gravitational energy released in the accretion process rather than nuclear burning, the mass of Bm could be less than \ 13~\ms, and the disk would be almost edge-on.  
Inside the disk radius of $\approx$0\farcs15, the 1.2-mm continuum flux is \ 42~mJy, corresponding to \ $\approx$1~\ms\  (making the same assumptions as in Sect.~\ref{res_stru}). Since this value is much less than the YSO mass $\approx$10~\ms, our choice of fitting a Keplerian velocity profile appears to be well justified \textit{a posteriori}. 

\begin{figure*}
\centering
\includegraphics[width=0.73\textwidth]{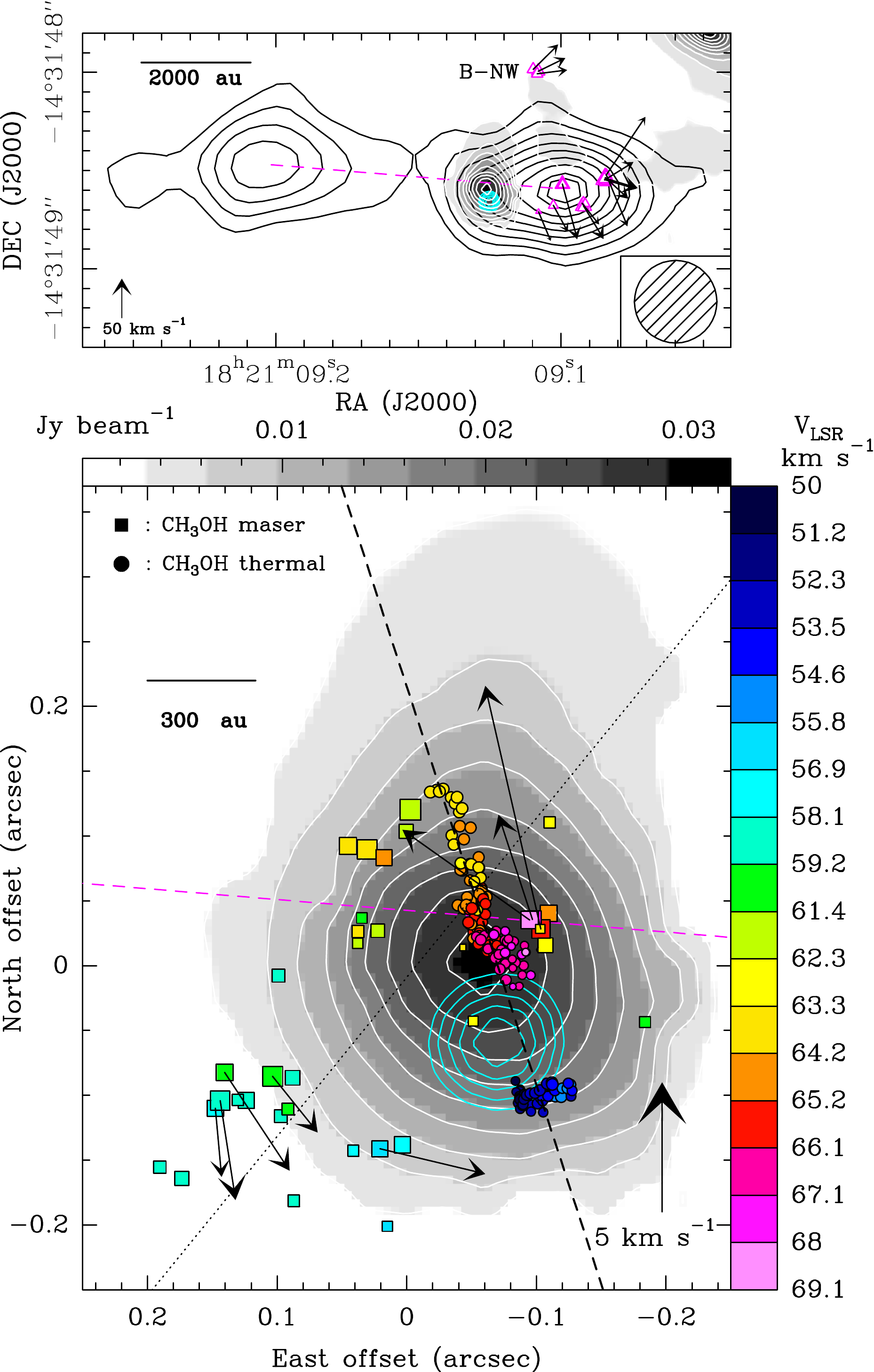}
\caption{{\em Lower~panel:}~Colored dots represent positions and \Vlsr\ of the thermal CH$_3$OH emission as in Fig.~\ref{peak_disk}, with the black dashed line marking the major axis of the disk around the high-mass YSO Bm. The grayscale map and the white contours reproduce the ALMA 1.2-mm continuum. The intensity scale for the map is shown on the top, and the plotted contours are the same as in Fig.~\ref{cont}. The JVLA 22-GHz continuum is shown with cyan contours (same levels as in Fig.~\ref{FM_CH3OH}). Colored squares denote positions and \Vlsr\ of the 6.7-GHz CH$_3$OH masers \citep{San10a}. The maser proper motions are represented by black arrows, with the amplitude scale given in the bottom-right corner. The black dotted line marks the major axis of the distribution of the CH$_3$OH masers. The dashed magenta line has the same meaning as in the upper panel. \ {\em Upper~panel:}~The black contours represent the JVLA 6-GHz continuum with levels from 20 to 90\%, in steps of 10\%, and 95\% of \ 0.17~m\Jyb. The dashed magenta line connects the peak emission of the eastern and western lobes of the radio jet. The grayscale map, and the white and cyan contours have the same meaning as in the lower panel. VLBI positions and proper motions of the water masers by \citet{San10a} are indicated with magenta triangles and black arrows, respectively. The amplitude scale for the proper motions is given in the bottom-left corner. The beam of the JVLA 6~GHz continuum image is reported in the bottom-right corner.}
\label{YSOpm}
\end{figure*}

Figure~\ref{YSOpm} (lower~panel) compares the kinematics of the disk, traced with thermal CH$_3$OH, with that of the 6.7-GHz CH$_3$OH masers observed with VLBI by \citet{San10a}. The 6.7-GHz masers trace a slightly elongated structure oriented along a \ PA~($\approx$~$-$40\degr) quite different from that (18\degr) of the YSO disk. The maser bipolar \Vlsr\ distribution (red-(blue-)shifted to NW(SE)) and the proper motion pattern were interpreted by   \citet{San10a} in terms of rotation, and the 6.7-GHz masers were the first indication of the existence of a rotating disk around Bm. However, the new ALMA data indicate now that the 6.7-GHz masers are not tracing the disk midplane, as originally assumed. While the positions and \Vlsr\ of the NW red-shifted 6.7-GHz cluster are in reasonable agreement with the red-shifted side of the YSO disk, the SE blue-shifted cluster is placed \ 0\farcs1--0\farcs2 \ SE  of the disk midplane. Looking at Fig.~\ref{YSOpm},  the two groups of maser features with measured proper motions are projected on the sky close to the disk rotation axis (i.e., perpendicular to the disk major axis crossing Bm, at the peak of the compact 22-GHz continuum) and the proper motions are mainly directed perpendicular to this axis. These findings still suggest that the 6.7-GHz masers are mainly tracing rotation about Bm, although they may be partly offset from the disk midplane.

%

%

\section{The YSO, the disk, and the jet}
\label{YDJ}

In this section, we discuss in more detail the interactions among the YSO, the disk, and the jet.

\subsection{Jet structure and orientation}
\label{disk-jet}

Looking at Fig.~\ref{peak_disk}, one can note that, near Bm, the slightly resolved 13-GHz continuum is elongated to NW along a direction roughly perpendicular to the disk. As shown in Figs.~\ref{W-NW_K2}~and~\ref{W-NW_K7}, the axis of the funnel-like structure on top of the radio jet traced by the  \ CH$_3$CN \ emission at \Vlsr $\approx$59~\kms \ also has  a similar NW orientation. This structure can  be interpreted in terms of the jet cavity, and together with the shape of the 13-GHz emission near Bm, could mark the direction of ejection of the jet close to the YSO. At larger separation from the YSO, the 13-GHz continuum seems to bend and approach the E-W direction, which could suggest that precession or recollimation is taking place.

The nonthermal continuum emission in the two lobes of the radio jet, shown in the upper panel of Fig.~\ref{YSOpm}, emerges from shocks responsible for the acceleration of the electrons to relativistic velocities. These shocks originate near Bm where the jet impinges on high-density material and propagate away along the jet direction at a speed, assuming momentum conservation, of \ $V_{sh}$~=~$\sqrt{n_{jet}/n_{amb}}$~$V_{jet}$, where \ $n_{jet}$ \ and \ $V_{jet}$ \ are the jet density and ejection velocity, respectively, and \ $n_{amb}$ \ is the ambient density \citep[see, e.g.,][]{Mas93}. 
Looking at the upper panel of Fig.~\ref{YSOpm}, one can note that the western lobe is significantly closer to Bm, which could indicate that the nonthermal shock has moved more slowly towards W because of the higher ambient density, as already discussed in Sect.~\ref{jet-mol}. From the JVLA 6-GHz map we derive that
the eastern lobe is about three times more distant from Bm than the western lobe, which, assuming the jet is ejected symmetrically, agrees well with our previous estimate of a density contrast in the ambient gas by  a factor of approximately ten between the western and eastern side. We denote with \ $\Delta S^{W}$ \ and \ $\Delta S^{E}$ \ the separations (projected onto the line connecting the lobes -- see Fig.~\ref{YSOpm}) between the position of Bm  and the emission peaks of the western and eastern lobes, respectively, and with \ $V_{sh}^{W}$ \ and \ $V_{sh}^{E}$ \ the shock propagation velocities towards W and E, respectively. The time elapsed since the episode of ejection responsible for the present emission in the jet lobes is then \ $\Delta T$~=~$\Delta S^{W}$/$V_{sh}^{W}$~=~$\Delta S^{E}$/$V_{sh}^{E}$. 
From the 6-GHz map we obtain \ $\Delta S^{W}$~$\approx$~0\farcs33~or~$\approx$1200~au, and 
  \ $\Delta T$~$\approx$~6000/$V_{sh}^{W}$~yr, where \ $V_{sh}^{W}$ \ is expressed in  \kms.

\subsection{Proper motion and precession}
\label{pm-pre}

Comparing the upper to the lower panel of Fig.~\ref{YSOpm}, it is clear that the line connecting the 6-GHz lobes of the radio jet crosses source~B near the 1.2-mm peak but is displaced from Bm (pinpointed by the 22-GHz emission) by \ $\ga$100~mas towards N. This offset is significantly larger than the expected positional error of a few~tens of milliarcseconds  between the 6- and 22-GHz JVLA images\footnote{We note that the same phase calibrator, J1832-1035, was used for  both the 6-~and~22-GHz JVLA observations, which guarantees a good relative astrometry between the images at the two frequencies.} \citep{Mos13b,Mos16}. 
In Sect.~\ref{jet-dis} we noted the clear asymmetry between the NE red-shifted and SW blue-shifted sides of the YSO envelope--disk, with the NE side being much more extended than the SE one (see Figs.~\ref{FM_CH3OH}~and~\ref{YSOpm}). We argue now that both the displacement of Bm from the jet axis and the disk asymmetry could result from the relative motion of Bm with respect to the disk. We assume that Bm has moved from its original position,  at the disk center, towards S-SW, along the disk plane. 

What is causing the motion of Bm? One possibility is that Bm is orbiting
in a gravitationally bound, multiple stellar system. This would naturally
explain 
its motion relative to the disk. 
In this case, the observed disk would rotate around the multiple
system and mediate the accretion onto it.  Recent observations of
bound systems of low-mass YSOs surrounded by a disk show that the regular
distribution in mass and velocity of the disk is truncated at an inner
radius comparable to the separation among the most massive stellar members
\citep{Tob16,Vil18}. At radii comparable
to these stellar orbits, the disk fragments and gaps, rings,
and spirals appear, like those observed in protoplanetary disks. Looking at
Fig.~\ref{peak_disk}, one can note that the disk around Bm presents a regular (Keplerian)
velocity pattern until reaching the position of the YSO, marked by the
compact 22-GHz continuum. The mass responsible for the observed velocity
pattern appears to be confined inside a radius \ $\le$~200~au from Bm, which
implies that other (putative) YSOs embedded in the parental mm~source~B
are not dynamically relevant for the disk or the motion of Bm.

An alternative explanation for the motion of Bm
is that the YSO has been ejected from its original position at the disk center by dynamical interaction with one or more companions. 
The YSO moved through the disk plane until it reached the current position, traced by the 22-GHz continuum.
The distribution of the gas in the disk remained essentially unaffected by the movement of the YSO.
The fact that we see Keplerian rotation around Bm implies
that the material of the disk has been rotating fast enough to re-adjust its
velocity field to the new position of the YSO. This sets a lower limit to
the crossing time, $T_{cr}$, needed for the star to go from the original position to the
current position. For a stellar mass of 10~\ms\ and a radius of the Keplerian 
disk of 0\farcs15 (540 au), we derive a rotational period \ $T_{rot}$ = 4000~yr. 
Thus, the condition \ $T_{cr} \ge T_{rot} = 4000$~yr \ must be satisfied. Using this estimate of \ $T_{cr}$
and the distance of \ $\sim$0\farcs1 \ traveled by Bm,
the average speed of Bm must be \ $\le$1~\kms. This is in agreement
with the result that the LSR velocity of Bm of $V_{\star}~=~60.4~\pm~0.5$~\kms determined from the Keplerian fit (see Sect.~\ref{disk-pro})
corresponds to the systemic velocity of the molecular cloud, $V_{sys}~=~60~\pm~0.04$~\kms, obtained from single-dish observations \citep[see, e.g.,][]{Hil10}.

For a plausible mass accretion rate of  $\sim$10$^{-3}$~\msyr, $T_{cr}$ \ is sufficiently long
that most of the mass of Bm could be accreted after its ejection
from the original position.  In this scenario, the structure of
the disk has not been significantly affected by the motion of the YSO
through it, because the star has become sufficiently massive to dominate
the gravitational field of the region, only when it was already close to
the current position. 

We still have to explain why the jet is offset from the current position of the YSO.
One possibility is that the jet was ejected before the YSO left its original position
(marked by the intersection between the jet axis and the disk).
Following the analysis of Sect.~\ref{disk-jet} and assuming  \ $\Delta~T$~$\sim$~$T_{cr}$~$\ge$4000~yr, we derive that the velocities of the shocks must be  \ $\le$1~\kms. Such a low velocity, much less than the typical jet speeds $\ge$100~\kms\ \citep[see, e.g.,][]{Mas15}, implies a very high density ratio ($\sim$10$^4$) between the ambient gas and the jet. Another possibility is that the jet is ejected from the present location of Bm,
but is then recollimated on a larger scale along the axis of the magnetic field, which, on that scale, has not been affected by the movement of the YSO. This latter interpretation is supported by the morphology of the 13-GHz continuum and the funnel-like molecular emission observed near Bm, which, as noted in Sect.~\ref{disk-jet}, suggests that the jet emerges from Bm perpendicular to the Keplerian disk.


Looking at Fig.~\ref{YSOpm}, it is also interesting to note that the 1.2-mm peak is slightly offset from the jet axis in the direction of the 22-GHz continuum. This offset, 0\farcs03--0\farcs04, is comparable with the error in the relative astrometry between the JVLA and ALMA images, but it could have a physical explanation, too.  In fact, it could result from a combination of two components of dust emission with similar intensities: one characterized by high temperature and low column density at the position of the 22-GHz continuum, the other with low temperature and large column density at the disk center. In order to resolve these two putative components, we have redone the 1.2-mm continuum image uniformly weighting the visibilities, thus increasing the angular resolution from \ 0\farcs15 \ to \ 0\farcs12, but the image still presents a single peak. We conclude that, if two components are present, their separation must be \ $\le$0\farcs1.

\section{Conclusions}

\label{conc}

The main results of our Cycle~3 high-angular resolution (beam FWHM $\approx$~0\farcs15) ALMA observations towards the high-mass SFR G16.59$-$0.05 can  be resumed as follows. At the center of the main clump, the dust emission is resolved into four small (size $\sim$1\arcsec) mm~sources, among which the one (source~B) harboring the high-mass YSO (Bm) is the most prominent in molecular emission. Fitting unblended, optically thin lines of CH$_3$OCH$_3$ and CH$_3$OH, we determined temperatures for the gas inside the mm~sources in the range \ 42--131~K, and estimated masses of between \ 1 and 5~\ms. 

A well-defined \Vlsr\ gradient (prominent in CH$_3$OH, CH$_3$CN and C$^{34}$S) is traced in many high-density molecular tracers  at the position of the high-mass YSO Bm, and is oriented along a direction forming a large ($\approx$70\degr) angle with the radio jet previously revealed through sensitive JVLA observations. The P-V plots of this gradient present butterfly-like shapes, and the emission peaks of the molecular lines at high velocity draw linear patterns, indicating that we are observing rotation of the disk--envelope surrounding Bm.  The disk radius is  estimated to be \ $\approx$500~au, and the \Vlsr\ radial distribution is well reproduced by Keplerian rotation around a central mass of \ 10$\pm$2~\ms.


The position of Bm, pinpointed by the compact 22-GHz emission, is found to be offset by  \ $\ga$0\farcs1 \ S of the jet axis and the mm~peak of source~B. We explain these findings assuming that, following a multiple stellar event, Bm was ejected from the center of the parental mm~source and moved through the disk plane until it reached the current position. While the material of the disk had enough time to re-adjust its velocity field to the new position of the YSO, the distribution of matter and magnetic field, responsible for collimating the jet, remained basically unaffected by the movement of the YSO. 

\begin{acknowledgements}
V.M.R. is funded by the European Union's Horizon 2020 research and innovation programme under the Marie Sk\l{}odowska-Curie grant agreement No 664931. 
We thank the referee Todd Hunter for useful comments, which improved the paper. This paper makes use of the following ALMA data: ADS/JAO.ALMA\#2015.1.00408.S. ALMA is a partnership of ESO (representing its member states), NSF (USA) and NINS (Japan), together with NRC (Canada), MOST and ASIAA (Taiwan), and KASI (Republic of Korea), in cooperation with the Republic of Chile. The Joint ALMA Observatory is operated by ESO, AUI/NRAO and NAOJ.     
\end{acknowledgements}

%
%

\bibliographystyle{aa} 
\bibliography{biblio} 

\begin{thebibliography}{36}
\expandafter\ifx\csname natexlab\endcsname\relax\def\natexlab#1{#1}\fi

\bibitem[{{Artur de la Villarmois} {et~al.}(2018){Artur de la Villarmois},
  {Kristensen}, {J{\o}rgensen}, {Bergin}, {Brinch}, {Frimann}, {Harsono},
  {Sakai}, \& {Yamamoto}}]{Vil18}
{Artur de la Villarmois}, E., {Kristensen}, L.~E., {J{\o}rgensen}, J.~K.,
  {et~al.} 2018, \aap, 614, A26

\bibitem[{{Beuther} {et~al.}(2002){Beuther}, {Schilke}, {Menten}, {Motte},
  {Sridharan}, \& {Wyrowski}}]{Beu02d}
{Beuther}, H., {Schilke}, P., {Menten}, K.~M., {et~al.} 2002, \apj, 566, 945

\bibitem[{{Beuther} \& {Shepherd}(2005)}]{Beu05}
{Beuther}, H. \& {Shepherd}, D. 2005, in Cores to Clusters: Star Formation with
  Next Generation Telescopes, ed. M.~S.~N. {Kumar}, M.~{Tafalla}, \&
  P.~{Caselli}, 105--119

\bibitem[{{Beuther} {et~al.}(2006){Beuther}, {Zhang}, {Sridharan}, {Lee}, \&
  {Zapata}}]{Beu06}
{Beuther}, H., {Zhang}, Q., {Sridharan}, T.~K., {Lee}, C.-F., \& {Zapata},
  L.~A. 2006, \aap, 454, 221

\bibitem[{{Bonato} {et~al.}(2018){Bonato}, {Liuzzo}, {Giannetti}, {Massardi},
  {De Zotti}, {Burkutean}, {Galluzzi}, {Negrello}, {Baronchelli}, {Brand},
  {Zwaan}, {Rygl}, {Marchili}, {Klitsch}, \& {Oteo}}]{Bon18}
{Bonato}, M., {Liuzzo}, E., {Giannetti}, A., {et~al.} 2018, \mnras, 478, 1512

\bibitem[{{Briggs}(1995)}]{Bri95}
{Briggs}, D.~S. 1995, in Bulletin of the American Astronomical Society,
  Vol.~27, American Astronomical Society Meeting Abstracts, 1444

\bibitem[{{Carpenter} {et~al.}(1997){Carpenter}, {Meyer}, {Dougados}, {Strom},
  \& {Hillenbrand}}]{Car97}
{Carpenter}, J.~M., {Meyer}, M.~R., {Dougados}, C., {Strom}, S.~E., \&
  {Hillenbrand}, L.~A. 1997, \aj, 114, 198

\bibitem[{{Davies} {et~al.}(2011){Davies}, {Hoare}, {Lumsden}, {Hosokawa},
  {Oudmaijer}, {Urquhart}, {Mottram}, \& {Stead}}]{Dav11}
{Davies}, B., {Hoare}, M.~G., {Lumsden}, S.~L., {et~al.} 2011, \mnras, 416, 972

\bibitem[{{Farias} \& {Tan}(2018)}]{Far18}
{Farias}, J.~P. \& {Tan}, J.~C. 2018, \aap, 612, L7

\bibitem[{{Furuya} {et~al.}(2008){Furuya}, {Cesaroni}, {Takahashi}, {Codella},
  {Momose}, \& {Beltr{\'a}n}}]{Fur08}
{Furuya}, R.~S., {Cesaroni}, R., {Takahashi}, S., {et~al.} 2008, \apj, 673, 363

\bibitem[{{Ginsburg} {et~al.}(2018){Ginsburg}, {Bally}, {Goddi}, {Plambeck}, \&
  {Wright}}]{Gin18}
{Ginsburg}, A., {Bally}, J., {Goddi}, C., {Plambeck}, R., \& {Wright}, M. 2018,
  \apj, 860, 119

\bibitem[{{Hill} {et~al.}(2010){Hill}, {Longmore}, {Pinte}, {Cunningham},
  {Burton}, \& {Minier}}]{Hil10}
{Hill}, T., {Longmore}, S.~N., {Pinte}, C., {et~al.} 2010, \mnras, 402, 2682

\bibitem[{{Hillenbrand} \& {Hartmann}(1998)}]{Hil98}
{Hillenbrand}, L.~A. \& {Hartmann}, L.~W. 1998, \apj, 492, 540

\bibitem[{{Kuiper} \& {Hosokawa}(2018)}]{Kui18}
{Kuiper}, R. \& {Hosokawa}, T. 2018, \aap, 616, A101

\bibitem[{{Lampton} {et~al.}(1976){Lampton}, {Margon}, \& {Bowyer}}]{Lam76}
{Lampton}, M., {Margon}, B., \& {Bowyer}, S. 1976, \apj, 208, 177

\bibitem[{{Lee} {et~al.}(2017){Lee}, {Ho}, {Li}, {Hirano}, {Zhang}, \&
  {Shang}}]{Lee17}
{Lee}, C.-F., {Ho}, P.~T.~P., {Li}, Z.-Y., {et~al.} 2017, Nature Astronomy, 1,
  0152

\bibitem[{{Masqu{\'e}} {et~al.}(2015){Masqu{\'e}}, {Rodr{\'{\i}}guez},
  {Araudo}, {Estalella}, {Carrasco-Gonz{\'a}lez}, {Anglada}, {Girart}, \&
  {Osorio}}]{Mas15}
{Masqu{\'e}}, J.~M., {Rodr{\'{\i}}guez}, L.~F., {Araudo}, A., {et~al.} 2015,
  \apj, 814, 44

\bibitem[{{Masson} \& {Chernin}(1993)}]{Mas93}
{Masson}, C.~R. \& {Chernin}, L.~M. 1993, \apj, 414, 230

\bibitem[{{Matsumoto} {et~al.}(2017){Matsumoto}, {Machida}, \&
  {Inutsuka}}]{Mat17}
{Matsumoto}, T., {Machida}, M.~N., \& {Inutsuka}, S.-i. 2017, \apj, 839, 69

\bibitem[{{McMullin} {et~al.}(2007){McMullin}, {Waters}, {Schiebel}, {Young},
  \& {Golap}}]{McM07}
{McMullin}, J.~P., {Waters}, B., {Schiebel}, D., {Young}, W., \& {Golap}, K.
  2007, in Astronomical Society of the Pacific Conference Series, Vol. 376,
  Astronomical Data Analysis Software and Systems XVI, ed. R.~A. {Shaw},
  F.~{Hill}, \& D.~J. {Bell}, 127

\bibitem[{{Moscadelli} {et~al.}(2013){Moscadelli}, {Cesaroni},
  {S{\'a}nchez-Monge}, {Goddi}, {Furuya}, {Sanna}, \& {Pestalozzi}}]{Mos13b}
{Moscadelli}, L., {Cesaroni}, R., {S{\'a}nchez-Monge}, {\'A}., {et~al.} 2013,
  \aap, 558, A145

\bibitem[{{Moscadelli} {et~al.}(2016){Moscadelli}, {S{\'a}nchez-Monge},
  {Goddi}, {Li}, {Sanna}, {Cesaroni}, {Pestalozzi}, {Molinari}, \&
  {Reid}}]{Mos16}
{Moscadelli}, L., {S{\'a}nchez-Monge}, {\'A}., {Goddi}, C., {et~al.} 2016,
  \aap, 585, A71

\bibitem[{{Ossenkopf} \& {Henning}(1994)}]{Oss94}
{Ossenkopf}, V. \& {Henning}, T. 1994, \aap, 291, 943

\bibitem[{{Peters} {et~al.}(2011){Peters}, {Banerjee}, {Klessen}, \& {Mac
  Low}}]{Pet11}
{Peters}, T., {Banerjee}, R., {Klessen}, R.~S., \& {Mac Low}, M.-M. 2011, \apj,
  729, 72

\bibitem[{Reid {et~al.}(1988)Reid, Schneps, Moran, Gwinn, Genzel, Downes, \&
  R{\"o}nn{\"a}ng}]{Rei88}
Reid, M.~J., Schneps, M.~H., Moran, J.~M., {et~al.} 1988, ApJ, 330, 809

\bibitem[{{Rivilla} {et~al.}(2016){Rivilla}, {Fontani}, {Beltr{\'a}n},
  {Vasyunin}, {Caselli}, {Mart{\'{\i}}n-Pintado}, \& {Cesaroni}}]{Riv16}
{Rivilla}, V.~M., {Fontani}, F., {Beltr{\'a}n}, M.~T., {et~al.} 2016, \apj,
  826, 161

\bibitem[{{Rodr{\'{\i}}guez-Kamenetzky}
  {et~al.}(2017){Rodr{\'{\i}}guez-Kamenetzky}, {Carrasco-Gonz{\'a}lez},
  {Araudo}, {Romero}, {Torrelles}, {Rodr{\'{\i}}guez}, {Anglada},
  {Mart{\'{\i}}}, {Perucho}, \& {Valotto}}]{Rod17}
{Rodr{\'{\i}}guez-Kamenetzky}, A., {Carrasco-Gonz{\'a}lez}, C., {Araudo}, A.,
  {et~al.} 2017, \apj, 851, 16

\bibitem[{{Sanchez-Monge} {et~al.}(2017){Sanchez-Monge}, {Schilke}, {Ginsburg},
  {Cesaroni}, \& {Schmiedeke}}]{Sanc17}
{Sanchez-Monge}, A., {Schilke}, P., {Ginsburg}, A., {Cesaroni}, R., \&
  {Schmiedeke}, A. 2017, ArXiv e-prints

\bibitem[{{Sanna} {et~al.}(2018{\natexlab{a}}){Sanna}, {Koelligan},
  {Moscadelli}, {Kuiper}, {Cesaroni}, {Pillai}, {Menten}, {Zhang}, {Garatti},
  {Goddi}, {Leurini}, \& {Carrasco-Gonzalez}}]{San18}
{Sanna}, A., {Koelligan}, A., {Moscadelli}, L., {et~al.} 2018{\natexlab{a}},
  ArXiv e-prints

\bibitem[{{Sanna} {et~al.}(2010){Sanna}, {Moscadelli}, {Cesaroni}, {Tarchi},
  {Furuya}, \& {Goddi}}]{San10a}
{Sanna}, A., {Moscadelli}, L., {Cesaroni}, R., {et~al.} 2010, \aap, 517, A71+

\bibitem[{{Sanna} {et~al.}(2018{\natexlab{b}}){Sanna}, {Moscadelli}, {Goddi},
  {Krishnan}, \& {Massi}}]{San18b}
{Sanna}, A., {Moscadelli}, L., {Goddi}, C., {Krishnan}, V., \& {Massi}, F.
  2018{\natexlab{b}}, ArXiv e-prints

\bibitem[{{Sato} {et~al.}(2014){Sato}, {Wu}, {Immer}, {Zhang}, {Sanna}, {Reid},
  {Dame}, {Brunthaler}, \& {Menten}}]{Sat14}
{Sato}, M., {Wu}, Y.~W., {Immer}, K., {et~al.} 2014, \apj, 793, 72

\bibitem[{{Tan} {et~al.}(2014){Tan}, {Beltr{\'a}n}, {Caselli}, {Fontani},
  {Fuente}, {Krumholz}, {McKee}, \& {Stolte}}]{Tan14}
{Tan}, J.~C., {Beltr{\'a}n}, M.~T., {Caselli}, P., {et~al.} 2014, Protostars
  and Planets VI, 149

\bibitem[{{Tobin} {et~al.}(2016){Tobin}, {Kratter}, {Persson}, {Looney},
  {Dunham}, {Segura-Cox}, {Li}, {Chandler}, {Sadavoy}, {Harris}, {Melis}, \&
  {P{\'e}rez}}]{Tob16}
{Tobin}, J.~J., {Kratter}, K.~M., {Persson}, M.~V., {et~al.} 2016, \nat, 538,
  483

\bibitem[{{Vaidya} {et~al.}(2011){Vaidya}, {Fendt}, {Beuther}, \&
  {Porth}}]{Vai11}
{Vaidya}, B., {Fendt}, C., {Beuther}, H., \& {Porth}, O. 2011, \apj, 742, 56

\bibitem[{{Zapata} {et~al.}(2006){Zapata}, {Rodr{\'{\i}}guez}, {Ho}, {Beuther},
  \& {Zhang}}]{Zap06}
{Zapata}, L.~A., {Rodr{\'{\i}}guez}, L.~F., {Ho}, P.~T.~P., {Beuther}, H., \&
  {Zhang}, Q. 2006, \aj, 131, 939

\end{thebibliography}

\end{document}